\documentclass[useAMS,usenatbib,usegraphicx]{mn2e} 

\usepackage{graphicx}
\usepackage{amssymb}
\usepackage{subfigure}
\usepackage{captcont}
\usepackage{multirow}
\usepackage{longtable}
\usepackage{comment}
\usepackage{url}
\usepackage{lscape}
\usepackage{booktabs,multirow}
\usepackage{times} 
\usepackage{txfonts}
\usepackage{aas_macros} 
\usepackage[usenames]{color}

\newcommand{\cd}{d$^{-1}$}
\newcommand{\kms}{km\,s$^{-1}$}

\title[HD 24355: A distorted roAp star in K2]{HD 24355 observed by the Kepler K2 mission: A rapidly oscillating Ap star pulsating in a distorted quadrupole mode\thanks{Based on service observations made with the WHT operated on the island of La Palma by the Isaac Newton Group in the Spanish Observatorio del Roque de los Muchachos of the Instituto de Astrof\'i­sica de Canarias.}}
\author[D. L. Holdsworth et al.]{Daniel L. Holdsworth$^{1,2}$\thanks{E-mail:dlholdsworth@uclan.ac.uk},
Donald W. Kurtz$^{1}$,
Barry Smalley$^{2}$,
Hideyuki Saio$^{3}$,
\newauthor Gerald Handler$^{4}$,
Simon J. Murphy$^{5}$,
and Holger Lehmann$^{6}$\\
$^{1}$ Jeremiah Horrocks Institute, University of Central Lancashire, Preston PR1 2HE, UK\\
$^{2}$ Astrophysics Group, Keele University, Staffordshire ST5 5BG, UK\\
$^{3}$ Astronomical Institute, School of Science, Tohoku University, Sendai 980-8578, Japan\\
$^{4}$ Copernicus Astronomical Center, Bartycka 18, 00-716, Warsaw, Poland\\
$^{5}$ Sydney Institute for Astronomy (SIfA), School of Physics, University of Sydney, NSW 2006, Australia\\
$^{6}$ Th\"{u}ringer Landessternwarte Tautenburg (TLS), Sternwarte 5, D-07778 Tautenburg, Germany\\
}

\begin{document}

\date{\today}

\pagerange{\pageref{firstpage}--\pageref{lastpage}} \pubyear{2016} 

\maketitle

\label{firstpage}

\begin{abstract}
We present an analysis of the first {\it Kepler} K2 mission observations of a rapidly oscillating Ap (roAp) star, HD\,24355 ($V=9.65$). The star was discovered in SuperWASP broadband photometry with a frequency of 224.31\,\cd\, (2596.18\,$\muup$Hz; $P = 6.4$\,min) and an amplitude of 1.51\,mmag, with later spectroscopic analysis of low-resolution spectra showing HD\,24355 to be an A5\,Vp\,SrEu star. The high precision K2 data allow us to identify 13 rotationally split sidelobes to the main pulsation frequency of HD\,24355. This number of sidelobes combined with an unusual rotational phase variation show this star to be the most distorted quadrupole roAp pulsator yet observed. In modelling this star, we are able to reproduce well the amplitude modulation of the pulsation, and find a close match to the unusual phase variations. We show this star to have a pulsation frequency higher than the critical cut-off frequency. This is currently the only roAp star observed with the {\it Kepler} spacecraft in Short Cadence mode that has a photometric amplitude detectable from the ground, thus allowing comparison between the mmag amplitude ground-based targets and the $\umu$mag spaced-based discoveries. No further pulsation modes are identified in the K2 data, showing this star to be a single-mode pulsator.
\end{abstract}


\section{Introduction}
\label{sec:intro}

The rapidly oscillating Ap (roAp) stars are a rare subclass of the chemically peculiar, magnetic, Ap stars. They show pulsations in the range of $6-23$\,min with amplitudes up to 18\,mmag in Johnson $B$ \citep{holdsworth15}, and are found at the base of the classical instability strip on the Hertzsprung-Russell (HR) diagram, from the zero-age main-sequence to the terminal-age main-sequence in luminosity. Since their discovery by \citet{kurtz82}, only 61 of these objects have been identified (see \citealt{smalley15} for a catalogue). The pulsations are high-overtone pressure modes (p~modes) thought to be driven by the $\kappa$-mechanism acting in the H\,{\sc{i}} ionisation zone \citep{balmforth01}. However, \citet{cunha13} have shown that turbulent pressure in the convective zone may excite some of the modes seen in a selection of roAp stars.

The pulsation axis of these stars is inclined to the rotation axis, and closely aligned with the magnetic one, leading to the oblique pulsator model \citep{kurtz82,ss85a,ss85b,dg85,st93,ts94,ts95,bigot02,bigot11}. Oblique pulsation allows the pulsation modes to be viewed from varying aspects over the rotation cycle of the star, giving constraints on the pulsation geometry that are not available for any other type of pulsating star (other than the Sun, which is uniquely resolved).

The mean magnetic field modulus in Ap stars are strong, of the order of a few kG to 34\,kG \citep{babcock60}. A strong magnetic field suppresses convection and provides stability to allow radiative levitation of some elements -- most spectacularly singly and doubly ionised rare earth elements --  producing a stratified atmosphere with surface inhomogeneities. These inhomogeneities, or spots, are long lasting (decades in many known cases) on the surface of Ap stars, thus allowing for an accurate determination of the rotation period of the star. In the spots rare earth elements such as La, Ce, Pr, Nd, Sm, Eu, Gd, Tb, Dy and Ho, may be overabundant by up to a million times the solar value, leading to spectral line strength variations over the rotation period \citep[e.g.][]{lueftinger10}. Because of the complex atmospheres of the Ap stars, the roAp stars provide the best laboratory, beyond the Sun, to study the interactions between pulsations, rotation, and chemical abundances in the presence of magnetic fields.

Early photometric campaigns targeted known Ap stars in the search for oscillations \citep[e.g.][]{martinez91,martinez94}, with later studies using high-resolution spectroscopy to detect line profile variations in Ap stars caused by pulsational velocity shifts \citep[e.g.][]{savanov99,koch01,hatzes04,mkr08,elkin10,elkin11,kochukhov13}. Most recently, the use of the SuperWASP (Wide Angle Search for Planets) ground-based photometric survey led to the identification of 11 roAp stars \citep{holdsworth14a,holdsworth15}.

With the launch of the {\it Kepler} space telescope, the ability to probe to $\muup$mag precision has enabled the detection of four roAp stars with amplitudes below the ground-based detection limit: KIC\,8677585 was a known A5p star observed during the 10-d commissioning phase of the {\it Kepler} mission and was shown to pulsate by \citet{balona11a}; KIC\,10483436 \citep{balona11b} and KIC\,10195926 \citep{kurtz11} were identified as roAp stars through analysis of their light curves and subsequent spectra; KIC\,4768731 was identified as an Ap star by \citet{niemczura15} and was later shown to be a roAp star \citep{smalley15}. Finally, {\it Kepler} observations also allowed the analysis of one roAp star, KIC\,7582608, identified in the SuperWASP survey with an amplitude of 1.45\,mmag \citep{holdsworth14b}, albeit in the super-Nyquist regime \citep{murphy13}.

There is an obvious difference between the roAp stars discovered with ground-based photometry, and those first detected by space-based observations: the amplitudes of the pulsations in the ground-based discoveries are generally in the range $0.5 - 10$\,mmag, whereas the {\it Kepler} observations did not detect variations above 0.5\,mmag. Ground-based observations are usually made in the $B$-band where the pulsation amplitude is greatest for the roAp stars \citep{medupe98}, and {\it Kepler} observations are made in a broadband, essentially white, filter where the amplitudes may be a factor of two to three lower. This accounts for some of the difference between the two groups. Further to this, ground-based observations are limited due to sky transparency white noise in the frequency range in which roAp stars are observed to pulsate, thus affecting the minimum amplitude that can be detected, suggesting this may be a observational bias. However the question remains as to whether there is a fundamental difference between the two groups. Are the differences in amplitude solely due to selection effects, or do the stars show differences in their abundance anomalies, magnetic field strengths, ages or rotation rates? The observations at $\muup$mag precision of a roAp star discovered with ground-based photometry may begin to provide insight into this disparity.

Ground-based projects in the search for transiting exoplanets produce vast amounts of data on millions of stars (e.g. WASP, \citealt{pollacco06}; HATnet, \citealt{bakos04}; ASAS, \citealt{pojmanski97}; OGLE, \citealt{udalski92}; KELT, \citealt{pepper07}). These data can provide an excellent source of information on many thousands of variable stars. Indeed, many of these projects have been employed for that purpose \citep[e.g.][]{pepper08,hartman11,ulaczyk13,holdsworth14a}. The ability of these surveys to achieve mmag precision provides an extensive all-sky database in which to search for low-amplitude stellar variability, which can then be observed at much higher precision by space-based missions such as K2 \citep{howell14} and TESS \citep{ricker15}. 

One of the leading ground-based projects in the search for transiting exoplanets, which provides data for many millions of stars, is the SuperWASP survey. This project is a two-site, wide-field survey, with instruments located at the Observatorio del Roque de los Muchachos on La Palma (WASP-N) and the Sutherland Station of the South African Astronomical Observatory \citep[WASP-S][]{pollacco06}. Each instrument has a field-of-view of $\sim$64\,deg$^2$, with a pixel size of 13.7\,arcsec. Observations are made through a broadband filter covering a wavelength range of $4000-7000$\,\AA\, and consist of two consecutive $30$-s integrations at a given pointing, with pointings being revisited, typically, every $10$\,min. The data are reduced with a custom reduction pipeline \citep[see][]{pollacco06} resulting in a `WASP $V$' magnitude which is comparable to the Tycho-$2$ $V_t$\, passband.  Aperture photometry is performed at stellar positions provided by the USNO-B$1.0$\, input catalogue \citep{monet03} for stars in the magnitude range $5<V<15$.

As previously mentioned, one of the space-based missions which ground-based surveys can inform is the K2 mission. After the failure of a second of four reaction wheels, the {\it Kepler} spacecraft could no longer maintain its precise pointing towards the original single field-of-view. The loss of the reaction wheels now means that the spacecraft needs to account for the solar radiation pressure in a new way. This has been achieved by pointing the telescope in the orbital plane. This new configuration requires semi-regular $\sim$5.9-hr drift corrections, as well as momentum dumps through thruster firings every few days \citep{howell14}. To avoid sunlight entering the telescope, a new field is selected every approximately 80\,d. Such a procedure has led to the fields being labelled as `Campaigns'.

Due to the shutter-less exposures of the {\it Kepler} spacecraft, the pointing drift leads to changes in brightness of an observed star as it moves across the CCD. There exist several routines to perform corrections on a large scale \citep[e.g.][]{vanderburg14,handberg14} which aim to reduce the systematic noise in the light curve and resultant periodogram. With careful reduction of the raw K2 data, the science data gives a photometric precision within a factor of $3-4$ of that of {\it Kepler} for a 12$^{\rm th}$ magnitude G star \citep{howell14}.

The re-purposing of the {\it Kepler} spacecraft has opened up a host of new possibilities to observe variable stars at $\muup$mag precision. The changing field-of-view now allows the study of O-type stars \citep{buysschaert15}; the prospect of detecting semi-convection in B-type stars \citep{moravveji15}; observations of variable stars in nearby open clusters \citep{nardiello15}; observations of RR Lyrae stars beyond the Galaxy \citep{molnar15}; and now the first observations, in Short Cadence, of a classical, high-amplitude, roAp star.

HD\,24355 is a bright ($V=9.65$) rapidly oscillating Ap star, discovered by \citet{holdsworth14a}. Their data show a pulsation at 224.31\,\cd\, (2596.18\,$\muup$Hz; $P = 6.4$\,min) with an amplitude of 1.51\,mmag\, in the WASP broadband filter. In this paper, we present a detailed spectral classification followed by an in-depth discussion of the SuperWASP discovery data, alongside further ground-based observations. We then present an analysis of the K2 Campaign 4 data.

\section{Spectral Classification}
\label{sec:spec}

\begin{figure}
\includegraphics[width=\linewidth]{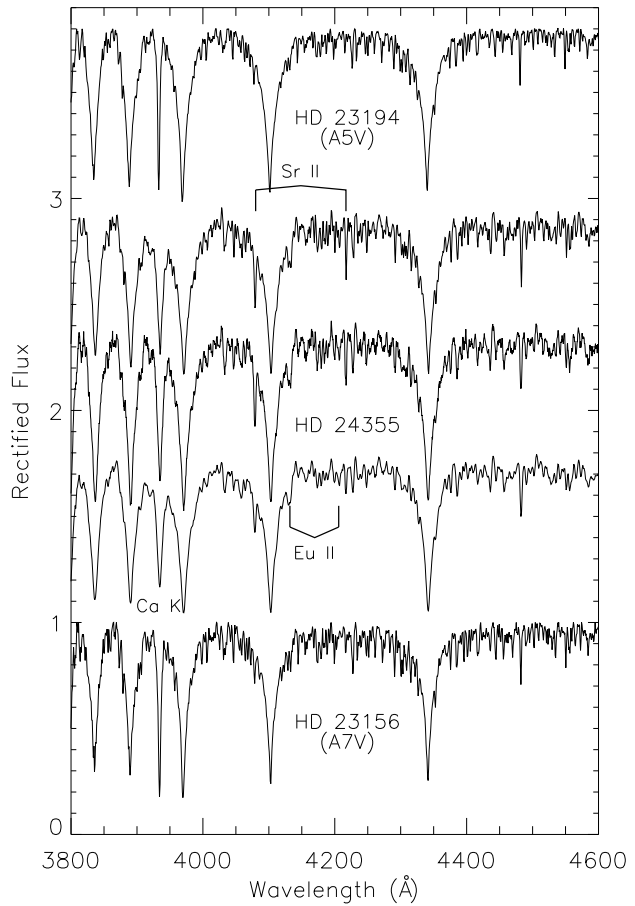}
\caption{Spectra of HD\,24355 (the three central spectra) compared with two MK standard stars of A5 and A7 type. Note the strong lines of Sr\,{\sc{ii}} at 4077 and 4216\,\AA\, and the enhanced lines of Eu\,{\sc{ii}} at 4130 and 4205\,\AA\, in the target star. The upper spectrum for HD\,24355 was obtained with the ANU/WiFes instrument, the second with the WHT/ISIS instrument, with the lower obtained with the LAMOST instrument. The spectra have been ordered in rotation phase and offset for clarity.}
\label{fig:spectra}
\end{figure}

We obtained two classification resolution spectra of HD\,24355 (Fig. \ref{fig:spectra}). The first, published in \citet{holdsworth14a}, was taken on 2012 October 25 with the Intermediate dispersion Spectrograph and Imaging System (ISIS) mounted on the 4.2-m William Herschel Telescope (WHT), at a resolution of $R\sim2000$. The second spectrum was obtained on 2014 November 15 using the Wide-Field Spectrograph (WiFeS) mounted on the 2.3-m  Australian National Observatory (ANU) telescope at Siding Spring Observatory at a resolution of $R\sim3000$. 

The ISIS data were reduced using the {\sc{figaro}} package \citep{shortridge04}, which is part of the {\sc{starlink}}\footnote{\url{http://starlink.jach.hawaii.edu/starlink}} suite of programmes. The WiFeS data were reduced with the PyWiFeS software package \citep{childress14}.

In addition to our data, HD\,24355 has been observed as part of the LAMOST campaign \citep{cui12,zhao12} and was made public as part of Data Release 1 \citep[DR1;][]{luo15}. The LAMOST spectrum was obtained at a resolution of $R\sim1800$, and is shown in Fig.\,\ref{fig:spectra} alongside the other spectra. Furthermore, in addition to classification spectra, we have obtained high-resolution, $R\sim58\,000$, spectra using the coud\'{e} \'{e}chelle spectrograph mounted on the 2.0-m telescope at the Th\"{u}ringer Landessternwarte (TLS) Tautenburg, Germany, which we discuss in section \ref{sec:High-Res}. Table\,\ref{tab:spec} gives details of all spectral observations. The phases have been calculated relative to the second pulsational maximum in the K2 data set, such that:
\begin{equation}
  \phi(E)=245\,7108.8867 + \mbox{27{\fd}9158}\times E,
  \label{equ:rot}
\end{equation}
where $E$ is the number of rotation cycles elapsed since the reference time. 

\begin{table*}
  \caption{Observation details for the all spectra. The S/N was determined using the  {\sc{der\_snr}} code of \citet{stoehr08}. The rotation phase has been calculated from equation\,(\ref{equ:rot}), as shown in the text.}
  
  \centering
  \label{tab:spec}
  \begin{tabular}{cccccc}
    \hline
    Instrument & BJD-245\,0000.0 & Exposure time & S/N & Resolution & Rotation\\
               &                 &   (s)         &     &            & phase   \\
    \hline
    ISIS  & 6225.6426     &  15           & 50  & 2\,000  &0.360 \\
    LAMOST& 6295.0420     &  2400         & 80  & 1\,800  &0.846 \\
    WiFeS & 6977.1331     &  300          & 30  & 3\,000  &0.280 \\
    TLS   & 7375.5077     &  2400         & 70  & 58\,000 &0.551 \\
    TLS   & 7380.4749     &  2400         & 65  & 58\,000 &0.729 \\
    \hline
    \end{tabular}
\end{table*}

Following the classification process described by \citet{gray09}, we see that the Balmer line types are close matches to the A5 and A7 MK standard stars\footnote{Spectra of the standard stars were obtained from R.O. Gray's website: \url{http://stellar.phys.appstate.edu/Standards/std1_8.html}} (HD\,23194 and HD\,23156, respectively) with many of the metal lines being well matched by the standard A5 star. There are, however, peculiarities which do not match the standard stars. Both spectra show strong enhancements of Sr\,{\sc{ii}} at 4077 and 4216\,\AA, and Eu\,{\sc{ii}} at 4130 and 4205\,\AA. While the Eu\,{\sc{ii}} line at 4130\,\AA\ can be blended with a Si\,{\sc{ii}} doublet at 4128 and 4131\,\AA, we see no sign of Si\,{\sc{ii}} at 3856 and 3862\,\AA, which are used to confirm the Si peculiarity in Ap stars. Lines of Cr\,{\sc{ii}} at 3866, 4111 and 4172\,\AA\ are not enhanced when compared to the MK standard stars, ruling out Cr peculiarities. With the obvious presence of Sr\,{\sc{ii}} and Eu\,{\sc{ii}} lines in the spectra, we conclude that HD\,24355 is an A5Vp SrEu star.

The Ca\,{\sc{ii}} K line is also abnormal in this star, a common trait in the Ap stars. The line is much broader and shallower than those of the standard stars. As the broadening is apparent in just this single line, we attribute the abnormality to chemical stratification in the atmosphere in the presence of a magnetic field, similar to that shown by \citet{smalley15}; comparison between a model atmosphere and a high-resolution spectrum is needed to confirm this. There is no clear difference between the line strengths of the peculiar lines between the two spectra, but given the relatively small shift in rotation phase between them, this is not unexpected.

We derive an effective temperature for HD\,24355 by fitting the Balmer lines of our spectra. Given their low resolution, we are unable to determine a $\log g$ of the star, so fix this value at $4.0$ which is a reasonable approximation for this exercise. From the WiFeS spectrum we derive $T_{\rm eff}=8200\pm250$\,K and from the ISIS data we find $T_{\rm eff}=8200\pm200$\,K. 

The LAMOST spectra are automatically analysed by instrument specific software to determine fundamental parameters of the star. For HD\,24355 the pipeline suggests $T_{\rm eff}=8638\pm100$\,K, $\log g=4.15\pm0.41$ (cgs) and ${\rm [Fe/H]} = 0.7\pm0.22$. However, when trying to fit these parameters to the spectrum, we find a large disparity. Given the resolution of the spectrum, we are unable to determine a revised $\log g$ (so we set this to $4.0$ as above). We find that $T_{\rm eff}=8150\pm250$\,K fits the Balmer lines much better than the higher value of the LAMOST estimate, and  $8150\pm250$\,K is in agreement with the other two spectra.

\section{Spectral Analysis}
\label{sec:High-Res}

To analyse HD\,24355  in detail, we obtained two high-resolution spectra with the coud\'{e} \'{e}chelle spectrograph mounted on the 2.0-m telescope at the Th\"{u}ringer Landessternwarte (TLS) Tautenburg, Germany. These spectra have a resolution of $R=58\,000$, and cover a wavelength range from $4720-7360$\,\AA. They were reduced using the standard ESO {\sc{midas}} packages. Information about the spectra is given in Table\,\ref{tab:spec}.

To determine the $T_{\rm eff}$ and $\log g$ of HD\,24355, we measured the equivalent widths of 60 Fe\,{\sc{i}} lines and 41 Fe\,{\sc{ii}} lines using {\sc{uclsyn}} \citep{smith88,smith92,uclsyn}. The effective temperature was found to be $8700\pm100$\,K by requiring no dependence of abundance with excitation potential. For the $\log g$ determination, we tested the ionisation balance between Fe\,{\sc{i}} and Fe\,{\sc{ii}}, finding $\log g=4.2\pm0.2$ (cgs).

The Fe abundance measured from the 101 lines is $\log A({\rm Fe})=8.3\pm0.3$, indicating an overabundance relative to solar of $[{\rm Fe}/{\rm H}]=0.8$\,dex, based on the solar chemical composition presented by \citet{asplund09}.

We also use the Fe lines to establish the microturbulence velocity of HD\,24355. We require no trend between the calculated abundances and their equivalent width when the correct value of microturbulence is chosen. We find a low value ($\sim0.0$\,\kms) of microturbulence provides us with no trend.

Finally, we determine that the $v\sin i$ of HD\,24355 is low by fitting lines with a series of synthetic spectra with varying $v\sin i$ values. We derive an upper limit of $3.5$\,\kms. Although this is below the quoted resolution of the spectrum, as lines sample more than one resolution element it is possible to attain a sub-element velocity measurement. We provide an upper limit as we have not accounted for the effects of magnetic broadening or macroturbulence.

As well as using the spectra to determine the $T_{\rm eff}$ of HD\,24355, we have performed spectral energy distribution (SED) fitting. We use literature photometry from 2MASS \citep{skrutskie06}, $B$ and $V$ magnitudes from \citet{hog97}, USNO-B1 $R$ magnitude \citep{monet03}, CMC14 $r'$  from \citet{evans02}, and the TASS $I$ magnitude \citep{droege06} to reconstruct the SED. There is no evidence in our high-resolution spectra of interstellar reddening along the line-of-sight to HD\,24355.

The stellar $T_{\rm eff}$ value was determined by fitting a [M/H]$\,= \,+0.5$  \citet{kurucz93} model to the SED. The model fluxes were convolved with photometric filter response functions. A weighted Levenberg-Marquardt non-linear least-squares fitting procedure was used to find the solution that minimised the difference between the observed and model fluxes. We used a $\log g = 4.0\pm0.5$ for the fit. The uncertainty in $T_{\rm eff}$ includes the formal least-squares error and adopted uncertainties in $\log g$ of $\pm0.5$ and [M/H] of $\pm0.5$ added in quadrature. As a result of the SED fitting, we derive a temperature of $7290\pm180$\,K.

It is clear that the methods used provide a wide range of $T_{\rm eff}$ values for HD\,24355 ($8638\pm100$\,K, LAMOST; $8200\pm200$\,K, Balmer fitting; $8700\pm100$\,K, Fe lines; $7100\pm120$\,K, SED). This is similar to the Ap star 53\,Cam \citep{kochukhov04}. In their analysis, they used different methods to determine the $T_{\rm eff}$ and $\log g$ of the star, and arrive at different values. They conclude that flux redistribution and line blanketing \citep{pyper83} and vertical gradients in elemental abundances \citep{wade03} to be the major causes in their discrepancies. These factors, as well as the use of inaccurate metallicities in the synthetic spectrum, may account for the disparities in our measurements. We also note that we do not take into account the effects of the magnetic field when we derive our $T_{\rm eff}$, which can affect the result.

In the case of HD\,24355, we have shown the LAMOST derived temperature to be an overestimate based on the Balmer lines. The temperature derived from the SED fitting is much below that derived from the Balmer lines and is perhaps under-sampled with the available photometry combined with flux redistribution due to line blanketing. Stratification is a well known phenomenon in Ap stars, and as such we suspect this is also the case here, where we have attempted to use the Fe lines to derive the Temperature. Therefore, based on the Balmer lines of both the High- and low-resolution spectra, we derive $T_{\rm eff}=8200\pm250$\,K for HD\,24355. At this temperature (and with the associated error), it is not possible to constrain the value of $\log g$, we therefore assume a value of $4.0\pm0.2$ (dex). 

\begin{figure}
  \includegraphics[width=\columnwidth,angle=180]{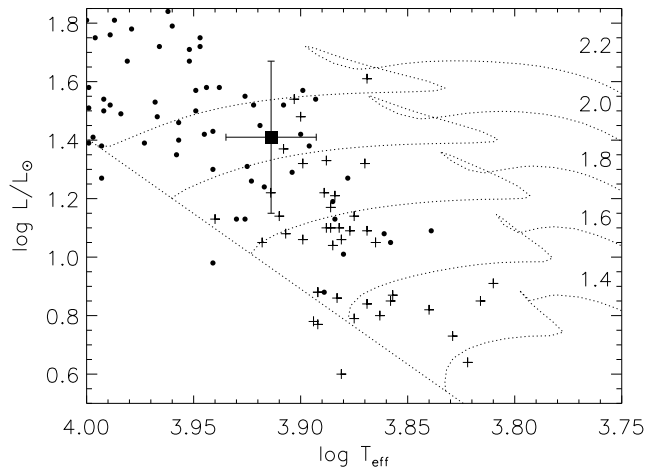}
  \caption{The position of HD\,24355 on the HR diagram, as derived from the spectra. The other roAp stars (pluses) and the noAp stars (circles) are also shown for context. The zero-age main-sequence and evolutionary tracks are from \citet{bertelli08}.}
    \label{fig:HRD}
\end{figure}

Using the relations of \citet{torres10}, and our values of $T_{\rm eff}=8200\pm200$, $\log g=4.0\pm0.2$ and $[{\rm Fe}/{\rm H}]=0.8\pm0.3$, we are able to derive the following parameters of the star: $M=2.40\pm0.38\,{\rm M_\odot}$, $R=2.53\pm0.43\,{\rm R_\odot}$. From these, we calculate the star's luminosity to be $\log(L/{\rm L_{\odot}})=1.41\pm0.26$. These parameters then allow us to plot HD\,24355 on a HR diagram, as shown in Fig.\,\ref{fig:HRD}, alongside other roAp stars and non-oscillating Ap (noAp) stars.

\subsection{Magnetic field}

Analyses of the two high-resolution spectra show there to be no magnetic splitting of spectral lines in this star. Therefore, to estimate the magnetic field strength we use the method of \cite{mathys92}. The ratio of strengths of Fe\,{\sc{ii}} 6147.7 and 6149.2\,\AA\, can be used as a proxy to estimate the mean magnetic field strength $\langle B\rangle$. Using the relative intensities of the lines, we measure a relative intensification of $\Delta W_{\lambda}/\overline{W}_{\lambda}$ to be $0.00\pm0.02$ and $0.08\pm0.02$ for the two spectra, respectively. Taking the mean of these measurements, and using the relation in \cite{mathys92}, this provides us with a mean magnetic field modulus of $2.12\pm1.44$\,kG. However, it is noted by \cite{mathys92} that the relation used cannot be extrapolated to weak fields, as is the case here, due to the different rate at which the two Fe\,{\sc{ii}} lines desaturate. We therefore treat this result only as an indication on the upper limit to the field strength.

\subsection{Identification of rare earth elements}

In Table\,\ref{tab:REE} we provide a first measurement of the rare earth element abundances in HD\,24355. As the two TLS spectra were obtained at similar rotation phases, we have co-added the two spectra and used the {\sc{uclsyn}} package \citep{smith88,smith92,uclsyn} to measured a selection of isolated rare earth element lines in the spectrum. As the table shows, there is a significant over-abundance of the rare earth elements in HD\,24355 when compared to the solar values, with some elements still enhanced when compared to other roAp stars \citep[e.g.][]{bruntt08,elkin11,smalley15}.

\begin{table}
  \caption{The abundances of the rare earth elements in HD\,24355. Abundances are given in the form $\log A({\rm El})=\log (N_{\rm EL}/N_{\rm H})+12.$ The second column, $n$, denotes the numbers of lines used to derive the abundance. The errors are the standard deviations of the $n$ measurements, and the solar abundances are from \citet{asplund09} for reference.}
 \centering
  \label{tab:REE}
  \begin{tabular}{lclc}
    \hline
    El & $n$ & $\log A({\rm El})$ & Solar\\
    \hline
    La & 5  & $3.14\pm0.34$ & 1.10 \\
    Ce & 4  & $3.47\pm0.33$ & 1.58 \\
    Pr & 8  & $3.17\pm0.66$ & 0.72 \\
    Nd & 16 & $4.02\pm0.97$ & 1.42 \\
    Sm & 3  & $3.10\pm0.53$ & 0.96 \\
    Eu & 5  & $2.78\pm0.56$ & 0.52 \\
    Gd & 6  & $3.47\pm0.26$ & 1.07 \\
    Tb & 3  & $3.42\pm0.10$ & 0.30 \\  
    Dy & 2  & $3.18\pm0.19$ & 1.10 \\
    Er & 3  & $3.13\pm0.29$ & 0.92 \\
    Tm & 1  & $3.3        $ & 0.10 \\
    Yb & 5  & $3.59\pm0.24$ & 0.84 \\
    Lu & 3  & $1.93\pm0.15$ & 0.10 \\
    \hline
  \end{tabular}
\end{table}
    
\section{Photometry}
\label{sec:phot}

The pulsation in HD\,24355 was discovered by \citet{holdsworth14a} after conducting a survey of the SuperWASP archive in a search for pulsating A stars. Here we provide a more detailed discussion of the discovery data, with the addition of further ground-based data obtained with the 0.75-m Automatic Photoelectric Telescope (APT) at Fairborn Observatory in Arizona, USA. Furthermore, we present the first Short Cadence (SC) data of a roAp star from the re-purposed {\it Kepler} space telescope, K2. HD\,24355 is the only Ap star observed in {\it{Kepler's}} SC mode whose rapid oscillation have high enough amplitude to be observed from the ground.

\subsection{Ground-based observations}
\label{sec:ground}

\subsubsection{SuperWASP}

SuperWASP observed HD\,24355 for three seasons, 2006, 2009 and 2010. The data were passed through a resistant mean algorithm to remove out-lying points to improve the quality of the periodogram; see \citet{holdsworth14a} for an example and details. After trimming the data, there were 7\,951 data points remaining.

\begin{figure}
  \includegraphics[width=\linewidth]{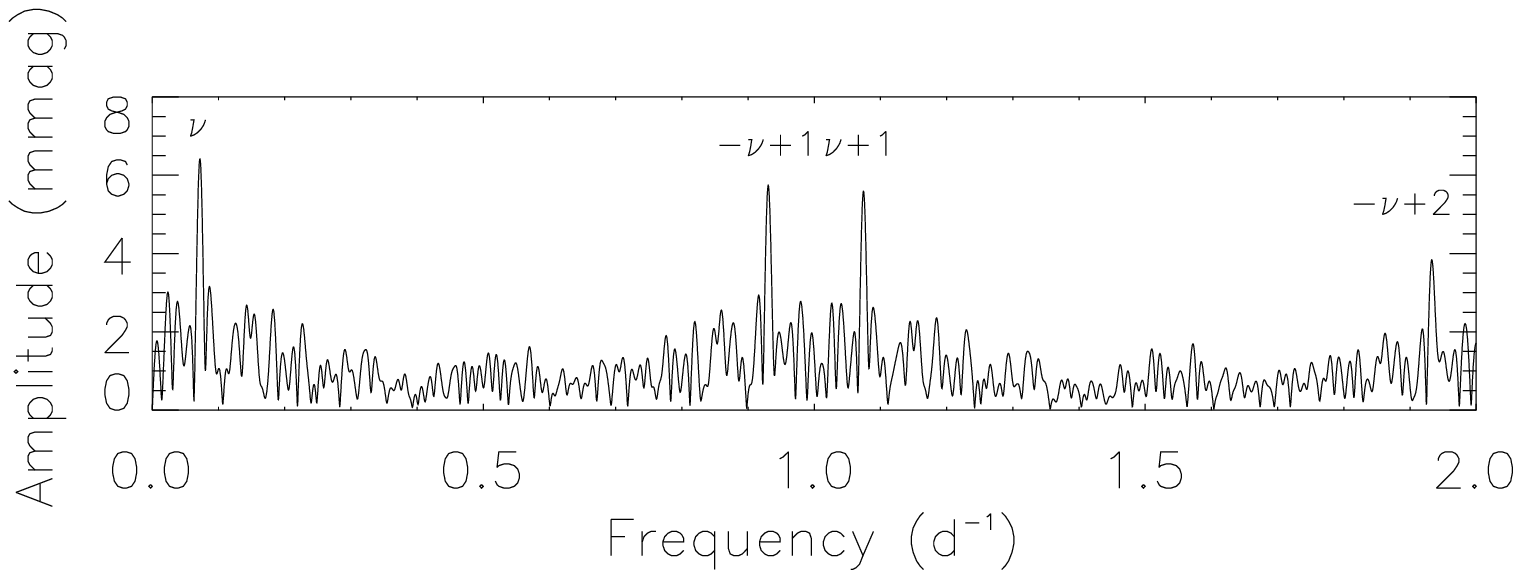}
  \includegraphics[width=\linewidth]{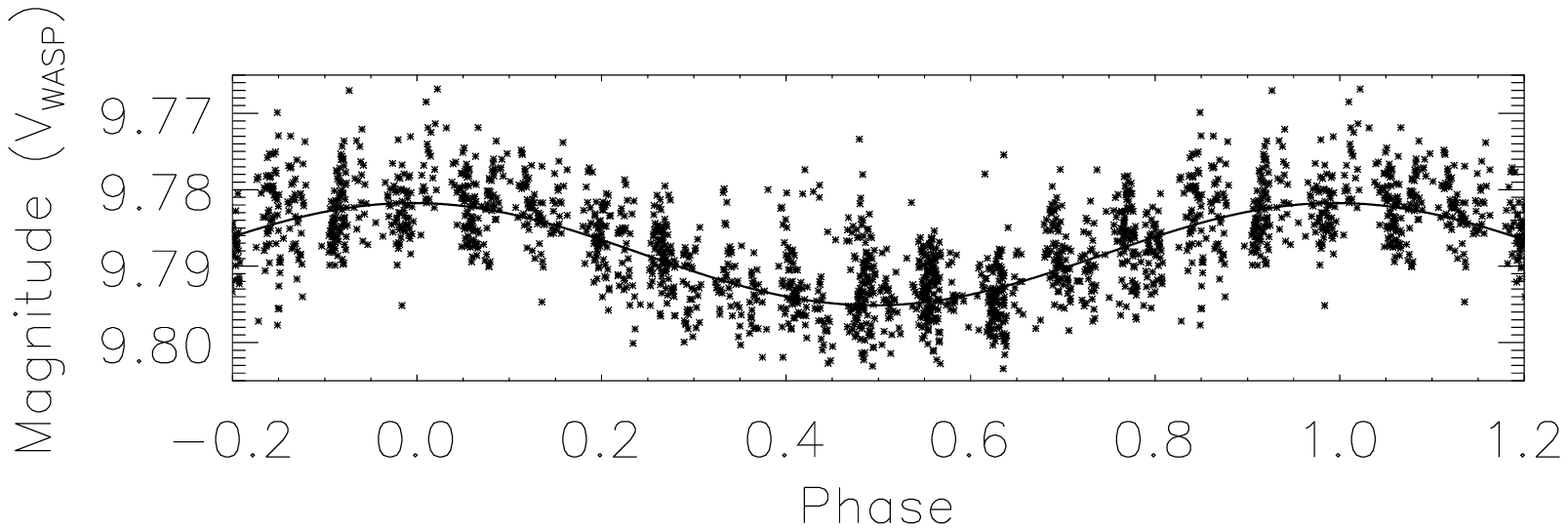}
  \caption{Top: low-frequency periodogram of the SuperWASP light curve from 2010 showing what was thought to be the rotation frequency of the star, as well as its positive and negative aliases. Bottom: SuperWASP phase folded light curve on the frequency determined in the upper panel, i.e. $\nu0.07165$\,\cd. The plot shows a clear single sinusoidal signature with a period of $1/\nu=13.9558$\,d. Analysis of the {\it Kepler} data in section\,\ref{sec:k2_rot} below shows that the true rotation period is twice this value. The data have been binned to 5:1. }
  \label{fig:lc}
\end{figure}

The light curve shows a clear signature at low frequency that is indicative of rotation (Fig.\,\ref{fig:lc}). The stable spots on Ap stars are usually aligned with the magnetic axis, which in turn is inclined to the rotation axis, leading to the rigid rotator model of \citet{stibbs50}, which manifests itself as a variation in brightness as the star rotates. As such, the rotation period of the star can be determined. Using the P{\sc{eriod}}04 program \citep{lenz05}, we detect a peak in the periodogram at a frequency of $0.0716439\pm0.0000057$\,\cd\, ($P=13.9579\pm0.0011$\,d), where the error is the analytical error given in P{\sc{eriod}}04, using the method of \citet{montgomery99}. However, when combining all seasons of WASP data to calculate the rotation period, there is evidence of a sub-harmonic to that frequency. We therefore we simultaneously calculate the peak at $0.0716$\,\cd\ and a frequency at half that value, namely $0.0358219$\,\cd. In doing so, we are able to a more precise  measure of the rotation frequency and its harmonic. We find, $\nu=0.0716439\pm0.0000063$\cd\ and $\nu/2=0.0358220\pm0.0000032$, corresponding to periods of $13.9579\pm0.0011$\,d and $27.9158\pm0.0025$\,d, respectively. In this ground-based survey data, differentiating between the rotation frequency and any harmonics can be problematic as low-amplitude peaks can be lost in the noise, especially in individual data sets. We therefore confirm the true rotation frequency with the K2 data, in section\,\ref{sec:K2}.

\begin{figure}
  \includegraphics[width=\linewidth]{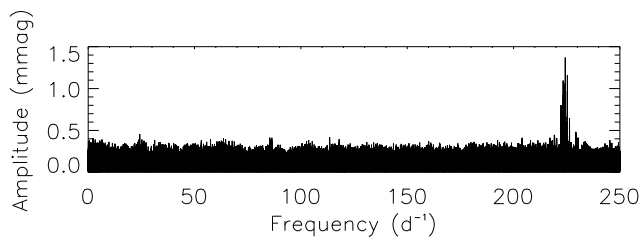}
  \caption{Periodogram of all seasons of WASP data. The data have been pre-whitened to 10\,\cd\, to the approximate noise level of the high-frequency range to remove the rotation signature and low-frequency noise. Note that the structure surrounding the pulsation signature is a result of daily aliases.}
  \label{fig:ft}
\end{figure}

The pulsation signature of HD\,24355 is apparent in all seasons of data, and is shown in Fig.\,\ref{fig:ft}. Each season of data has been pre-whitened, to 10\,\cd, to the approximate noise level of the high-frequency range to remove the rotation signature and the remaining low-frequency `red' noise after the data have been processed by the WASP pipeline \citep{smith06}. The pulsation is clearly seen at a frequency of 224.3071\,\cd\, with an amplitude of 1.51\,mmag in the WASP passband. The frequencies, amplitudes and phases of a non-linear least squares fit for each season are provided in Table \ref{tab:wasp-fits}.

\begin{table*}
    \caption{Details of the WASP observations, and the results of a non-linear least-squares fit of the pulsation frequency in each of the seasons. BJD is given as BJD-245\,0000.0, and the zero-point for the phases is taken to be the mid-point of the individual data sets.}
\label{tab:wasp-fits}
  \begin{tabular}{ccccccr}
    \hline
    Season & BJD & Length & Number of & Frequency & Amplitude & \multicolumn{1}{c}{Phase}\\
           & start & (d)  & points    &  (\cd)      & (mmag)    & \multicolumn{1}{c}{(rad)} \\
    \hline
    2006 & 3995.5894 & 129.7952 & 2083 & 224.3073$\pm$0.0004 & 1.89$\pm$0.18 &  2.391$\pm$0.089 \\
    2009 & 5148.4549 & 83.0317  & 1847 & 224.3048$\pm$0.0007 & 1.39$\pm$0.17 & -2.415$\pm$0.124 \\
    2010 & 5470.5701 & 125.9043 & 4021 & 224.3052$\pm$0.0004 & 1.55$\pm$0.12 & -2.201$\pm$0.077 \\
    \hline
  \end{tabular}
\end{table*}

\begin{figure}
  \includegraphics[width=\linewidth]{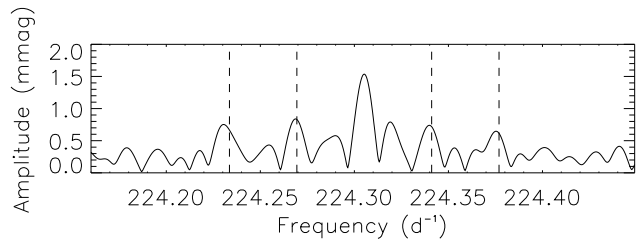}
  \caption{SuperWASP detections of sidelobes to the pulsation frequency in the 2010 season. The sidelobes are approximately at frequencies $\pm1$ and $\pm2$ times the rotational sub-harmonic frequency (as shown by the dashed lines). This suggests that the rotation frequency is 0.0358\,\cd\, i.e. a period of 27.9117\,d.}
  \label{fig:lobes}
\end{figure}

Due to the survey nature of this ground-based data set, not much further information can be extracted: the noise level in the high-frequency range is too great to discern, with confidence, sidelobes of the pulsation split by the rotation frequency. However, in the 2010 season of data there do appear to be peaks split from the pulsation by the sub-harmonic of the strongest low-frequency peak (Fig.\,\ref{fig:lobes}). This provides evidence that the rotation period is likely to be of the order 27\,d rather than 13\,d. Again, analysis of the K2 data in the following section resolves this disparity.

\subsubsection{0.75-m APT}

We observed HD\,24355 for three hours on each of three consecutive nights, 2013 November 6, 7, 8, with the 0.75-m Automatic Photoelectric Telescope (APT) installed at Fairborn Observatory in Arizona (Table\,\ref{tab:apt}), and later obtained data concurrent with the K2 observations (see section\,\ref{sec:APT-K2}). Observations were scheduled and taken remotely and consist of 20-s integrations through a Johnson $B$ filter with a photometric aperture of 30\,arcsec. No comparison stars were observed as the pulsation period is known and the fluctuations in the sky transparency occur at frequencies much lower than the pulsation and therefore do not interfere. Disregarding comparison stars enabled us to maximise our time on target and provide continuous observations per night.

The data were reduced and corrected for the dead-time of the detector, the sky values were fitted and subtracted from the target and the extinction corrected for using Bouguer's Law and a standard site extinction coefficient of 0.24\,mag/airmass for the $B$ filter. Finally, the data times were adjusted to BJD.

\begin{table*}
  \caption{Details of the APT observations, and the results of a non-linear least-squares fitting of the frequency and its harmonic. BJD is given as BJD-245\,0000.0, the rotation phase is calculated as in equation\,(\ref{equ:rot}), and the zero-point of the phases is the mid-point of the individual data sets. Note the increase in amplitude of $\nu_1$ over the three nights. } 
  \label{tab:apt}
  \begin{tabular}{ccccccccr}
    \hline
    Date & BJD   & Rotation & Length & Number of & Label & Frequency & Amplitude & \multicolumn{1}{c}{Phase} \\
    (UT) & start & phase    & (h)    & points    &       &(\cd)      &  (mmag)   & \multicolumn{1}{c}{(rad)} \\
    \hline
    2013-11-06 & 6602.8108 & 0.873 & 2.95 & 416 & $\nu_1$  & 224.04$\pm$0.22 & 3.52$\pm$0.17 & 1.586$\pm$0.050 \\
               &           &       &      &     & $2\nu_1$ & 448.06$\pm$0.64 & 1.21$\pm$0.17 & 2.847$\pm$0.144 \\
                                                                                                                 
    2013-11-07 & 6603.8021 & 0.907 & 3.02 & 416 & $\nu_1$  & 224.34$\pm$0.13 & 4.79$\pm$0.15 & -0.266$\pm$0.031 \\
               &           &       &      &     & $2\nu_1$ & 448.19$\pm$0.54 & 1.19$\pm$0.15 & -0.629$\pm$0.126 \\
                                                                                                                 
    2013-11-08 & 6604.8011 & 0.943 & 3.08 & 416 & $\nu_1$  & 224.41$\pm$0.11 & 5.84$\pm$0.16 &  2.290$\pm$0.026 \\
               &           &       &      &     & $2\nu_1$ & 448.40$\pm$0.41 & 1.64$\pm$0.16 & -1.797$\pm$0.095 \\
    \hline
  \end{tabular}
\end{table*}

Before analysing the pulsation in the APT data, we pre-whitened the light curves individually to a level of 0.6 mmag to a frequency of 24\,\cd\, with the aim of removing any remaining atmospheric effects not accounted for in the airmass correction and sky subtraction performed. As the pulsation frequency is significantly higher than this limit, this low-frequency filtering has no detrimental effect on our analysis.

\begin{figure}
  \includegraphics[width=\linewidth]{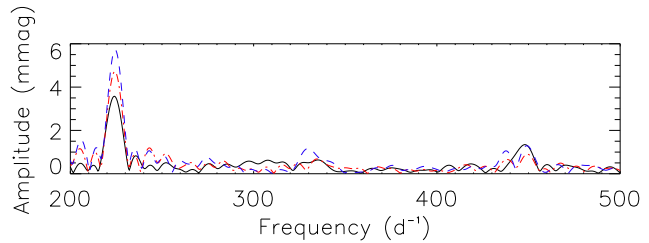}
  \caption{Periodograms of the three nights of APT data. There is a clear increase in the pulsation amplitude as the nights progress, from the solid black line, to the dot-dashed red line, to the dashed blue line. Also evident is the harmonic at twice the pulsation frequency.}
  \label{fig:APT}
\end{figure}

The periodograms for each night of data presented in Fig.\,\ref{fig:APT} show a clear increase in pulsation amplitude over the three nights that the target was observed. This is expected from the oblique pulsator model \citep{kurtz82,bigot11}, and is also observed in the K2 observations (see section\,\ref{sec:K2}). As the pulsation axis is often mis-aligned with the rotation axis in the roAp stars, the pulsation is viewed from varying aspects over the rotation period, giving rise to the amplitude variations we see in HD\,24355.

Also evident in the periodograms is the presence of the first harmonic of the principal pulsation (cf. Fig.\,\ref{fig:APT} and Table\,\ref{tab:apt}). This is often seen in roAp stars.

The amplitude of the pulsation increases over the three nights of observations (Table\,\ref{tab:apt}). Calculating the rotation phase at which the observations occur, using equation\,(\ref{equ:rot}), we are able to show they were obtained as the star was approaching maximum light in its rotation cycle, and maximum pulsation amplitude (see Section\,\ref{sec:freq_var}). The observations of increasing pulsation amplitude is therefore an expected outcome, as predicted by the oblique pulsator model.

\subsection{K2 observations}
\label{sec:K2}

\begin{figure}
  \includegraphics[width=\linewidth]{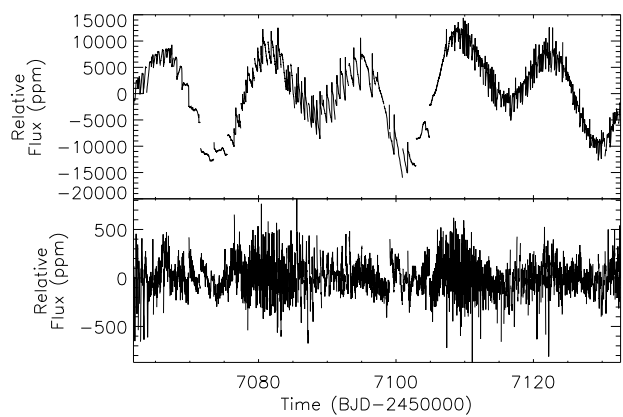}
  \caption{A comparison of the raw K2 Long Cadence light curve (top) for HD\,24355 and the K2P$^2$ pipeline light curve (bottom). Clearly, the low-frequency thruster firings are removed by the K2P$^2$, but so is the astrophysical signal of rotation. This is not surprising, since the K2P$^2$ pipeline was not designed to handle reduction of rotational variations on the time scale seen in HD\,24355. Note the extreme change in ordinate scale between the two panels.}
  \label{fig:K2_raw_K2P2}
\end{figure}

HD\,24355 was observed as part of Campaign 4 (proposal number GO4014) of the K2 mission. The observations covered a period of 70.90\,d. There are now available many community reduction pipelines in operation for K2 data, each resulting in slightly different light curves. This is particularly evident when dealing with the low-frequency rotational modulation which is seen in HD\,24355. The data reduction performed by the K2P$^2$ pipeline of \cite{Lund15} and \cite{handberg14} removes the low-frequency rotational variability as the code uses a running moving-median filter to remove the spacecraft thruster firings. Fig.\,\ref{fig:K2_raw_K2P2} shows a comparison between the raw K2 light curve and the K2P$^2$ light curve where clearly the rotational low-frequencies are removed. Currently, only long cadence data reduced with K2P$^2$ are available for HD\,24355. Therefore, rather than using a community provided reduction pipeline, we reduced the data manually using the PyKE tools provided by the K2 Guest Observer office \citep{still12}.

\subsubsection{Data reduction}

\begin{figure}
  \includegraphics[width=\columnwidth]{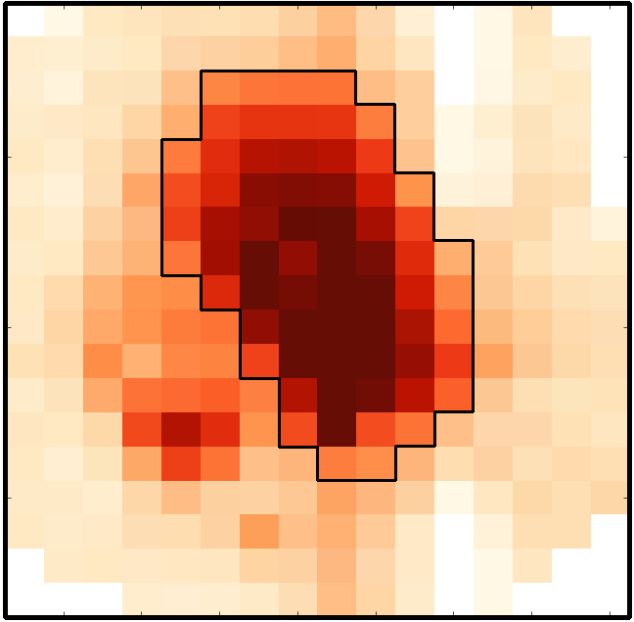}
  \caption{The pixel mask for HD\,24355. The entire area is that downloaded from the spacecraft, with the solid line being the mask used to extract the light curve in this work. Note the presence of V478 Tau in the lower left. The mask was chosen to minimise light from this source.}
  \label{fig:mask}
\end{figure}

The raw light curve was reduced with the PyKE tools. A pixel mask was chosen using the {\sc{kepmask}} tool, and is shown in Fig.\,\ref{fig:mask}. Care was taken to avoid the fainter object to the lower left of the target. This star, V478 Tau, is located 18.8\,arcsec from HD\,24355 and is a known $B=15.0$ flare star classified M3-M4 by \citet{parsamyan93}. We also note that the proper motion of V478 Tau differs from that of HD\,24355 indicating that the two stars are not physically associated.

\begin{figure*}
  \centering
  \begin{minipage}{\textwidth}
    \centering
    \includegraphics[width=\linewidth]{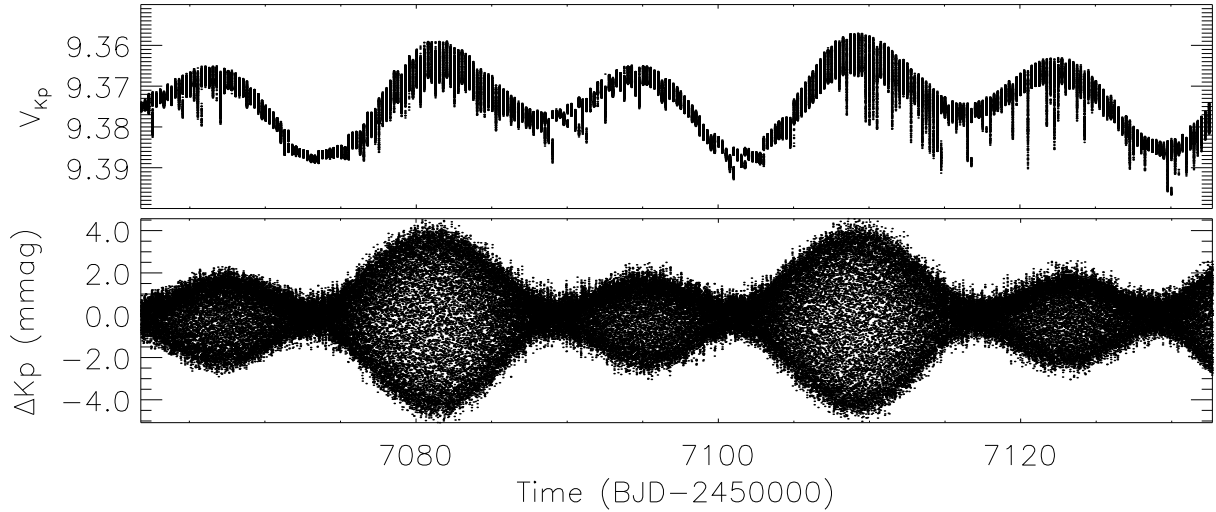}
    \caption{Top: full K2 light curve after data reduction and cleaning has been performed. Clearly evident is the rotational signal, as well as the spacecraft motions which have not been removed. Bottom: light curve of HD\,24355 after removal of the rotation signature, and cleaning of low-frequency peaks from the periodogram; these include the harmonics of the thruster frequency. The pulsation envelope is evident, with pulsational maximum occurring at the same time as rotation light maximum within the errors.}
    \label{fig:K2_lc}
  \end{minipage}
\end{figure*}

The light curve was extracted using {\sc{kepextract}}, which included a background subtraction, and then cleaned of obvious spurious points, resulting in 102\,917 data points with which we conducted the analysis. We did not attempt to remove the motion systematics from the data with the PyKE tools; in an attempt to do so, low-frequency stellar variability was also removed, just as with the K2P$^2$ pipeline. Due to the high amplitude of the rotation signature, we are still able to reliably extract the rotation period, despite the thruster artefacts. Further to this, the low-frequency systematics are far removed from the pulsation signal, so do not interfere with the pulsation analysis. The final light curve is shown in the top panel of Fig.\,\ref{fig:K2_lc}.

\subsubsection{Rotation signature}
\label{sec:k2_rot}

Despite the thruster firings still being present in the data, the rotation signature is clearly seen in Fig.\,\ref{fig:K2_lc}. The double-wave nature of the light curve suggests we see both magnetic poles of the star, where the chemically peculiar spots form. The K2 data allow us to unambiguously determine the rotation period of the star due to the highly stable nature of the spots coupled with the high photometric precision.

\begin{figure}
  \includegraphics[width=\columnwidth]{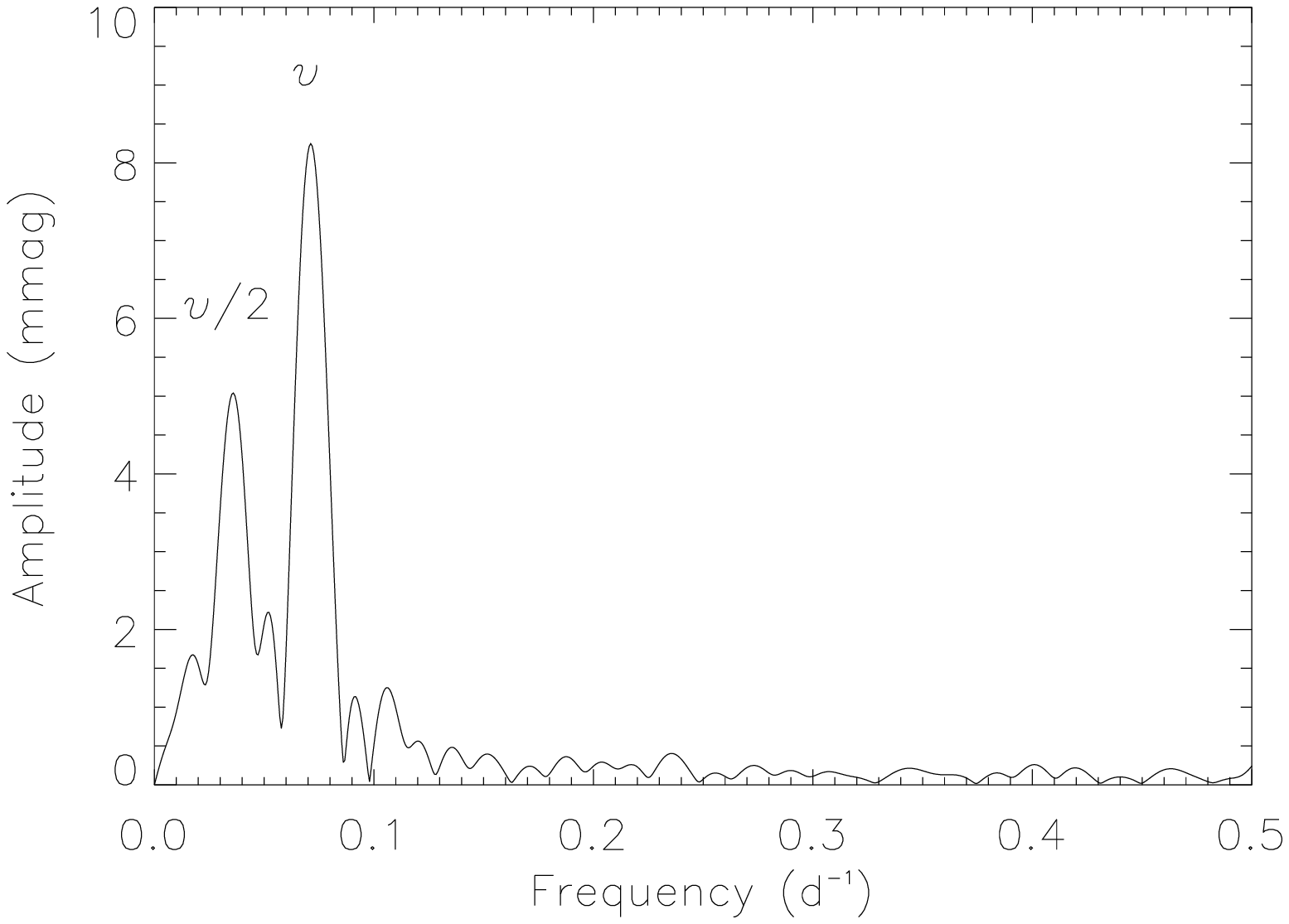}
  \includegraphics[width=\columnwidth]{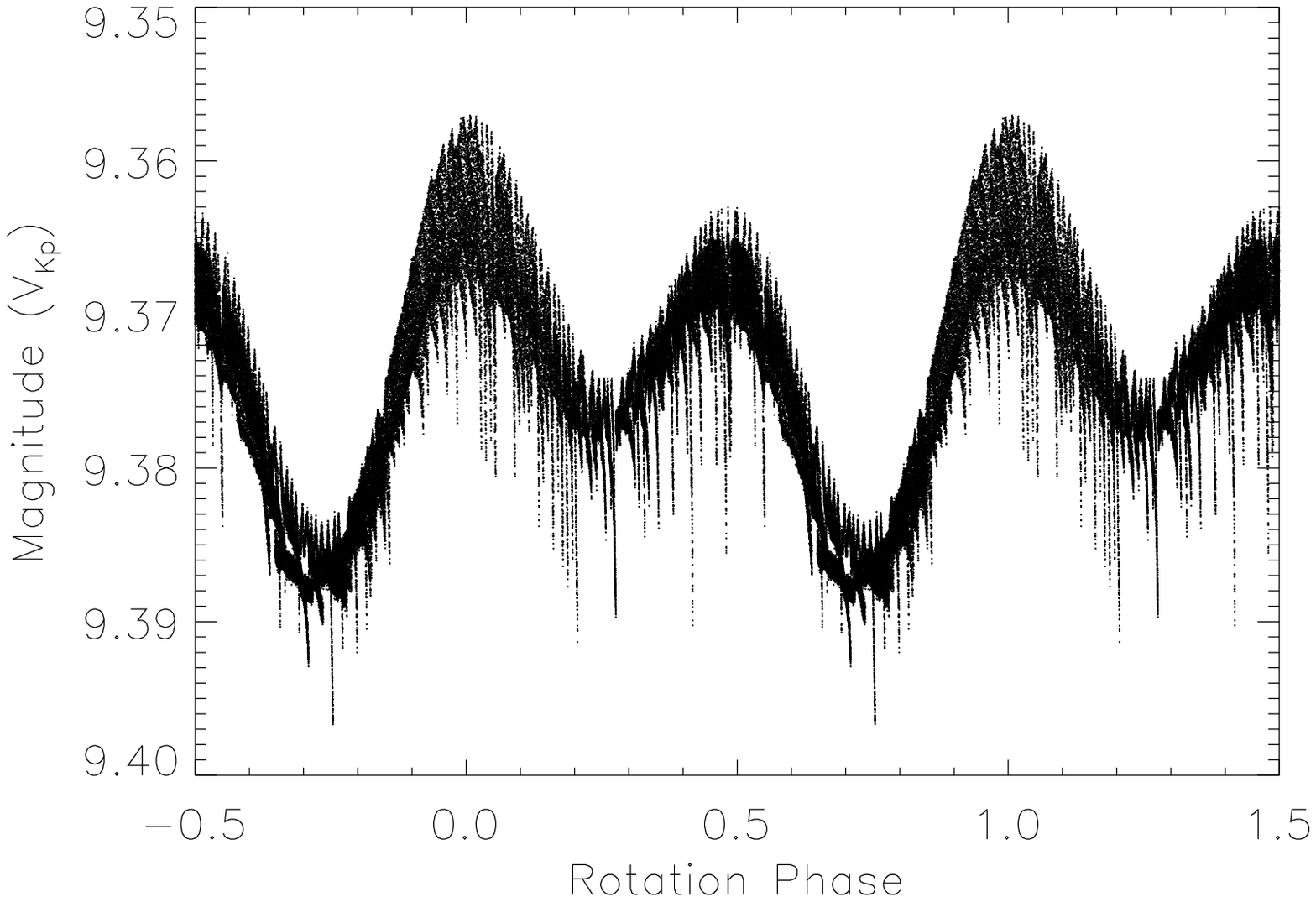}
  \caption{Top: periodogram of the K2 light curve, showing the strongest peak at $\nu$, as seen in WASP, and a sub-harmonic at $\nu/2$. Bottom: K2 phased light curve, phased on a period of 27.9166\,d, corresponding to the lower amplitude peak in the top panel. The phases have been calculated according to equation\,(\ref{equ:rot}).}
  \label{fig:phased_rot}
\end{figure}

Using P{\sc{eriod04}} we calculate an amplitude spectrum to 0.5\,\cd. The result shows two peaks, corresponding to the two consecutive dimmings in the light curve (Fig.\,\ref{fig:phased_rot}). The principal peak is found at a frequency of $0.071642 \pm 0.000011$\,\cd, with the secondary at $0.036121 \pm 0.000019$\,\cd, as seen in the WASP data. The errors presented here are the analytical errors calculated using P{\sc{eriod04}} following the method of \citet{montgomery99}. When phasing the data on the principal peak, it is clear that this is not the correct frequency which describes the modulation of the light curve, rather the lower amplitude peak provides a `clean' phased light curve (when ignoring the spacecraft motion). Fig.\,\ref{fig:phased_rot} shows the light curve phased on the lower frequency. This is clearly the correct rotation frequency. This then provides us with a model of a double-wave spot variation, and the knowledge that the two peaks in Fig.\,\ref{fig:phased_rot} are harmonics. Taking the highest amplitude harmonic, i.e. the peak at $0.071642$\,\cd, and dividing by two provides us with an accurate rotation frequency. We use the highest amplitude harmonic as its frequency can be determined to a higher precision. In following this procedure, we determine a rotation period of the star to be $27.9166\pm0.0043$\,d. This exercise has allowed us to determine the correct rotation frequency of HD\,24355, and remove the ambiguity seen in the WASP data. Due to the relatively short length of the K2 data (covering $\sim2.5$ rotation periods), we take the period derived from the WASP data (which covers $\sim57.3$ rotation periods) as the rotation period for HD\,24355, namely $P=27.9158\pm0.0025$\,d.

This result removes the ambiguity between the signals derived from the WASP data above. Due to the noise characteristics of the WASP data, especially at low frequencies, it was not possible to identify the correct rotation period from the light curve. We use the period derived from the K2 data throughout this work.

\subsubsection{Pulsation signature}

\begin{figure}
  \includegraphics[width=\columnwidth]{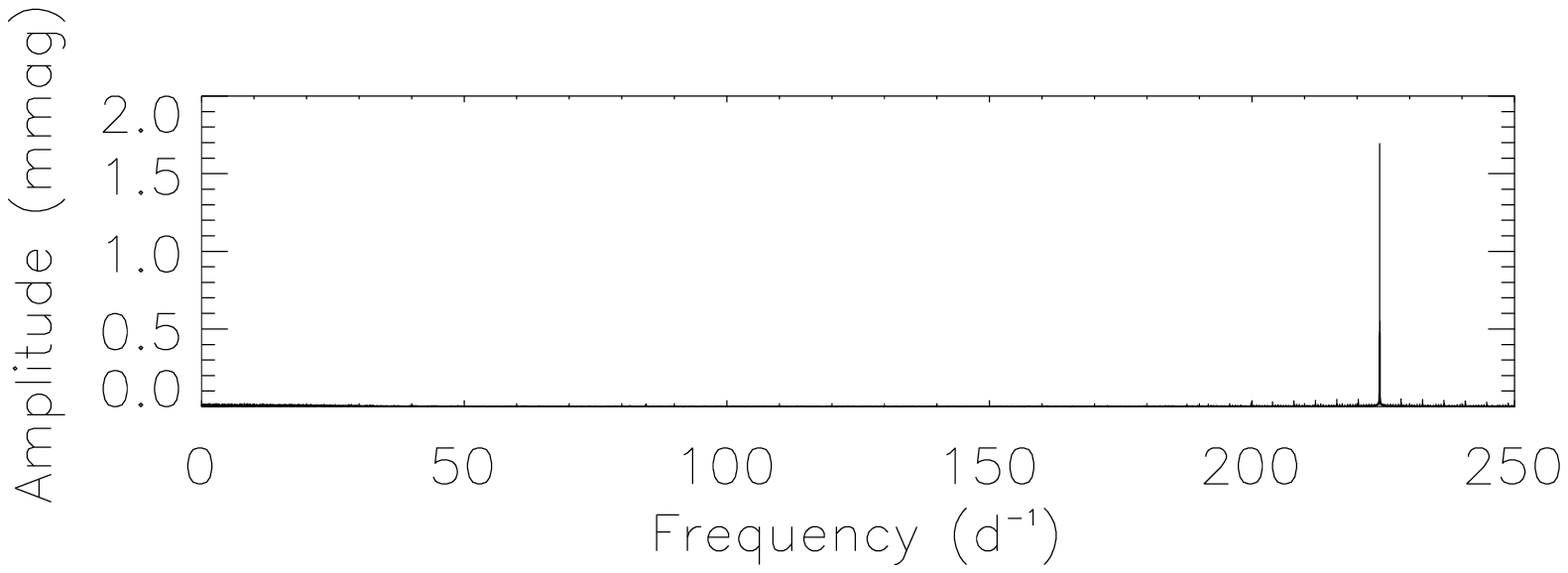}
  \caption{Periodogram of the reduced and cleaned K2 light curve. Clearly evident is the pulsation at 224.3\,\cd and its window pattern generated by the $\sim$6-hr thruster firings. There is still some low-frequency noise remaining in the data.}
  \label{fig:K2_all_ft}
\end{figure}

Clearly evident in the K2 periodogram of HD\,24355 is the pulsation detected by both the WASP and APT observations. To study the pulsation in detail, we remove the rotation signature from the light curve, and significant peaks in the low-frequency domain. This process results in the light curve shown in the bottom panel of Fig.\,\ref{fig:K2_lc}. This cleaned light curve consists of 101\,427 points. Fig.\,\ref{fig:K2_all_ft} shows a periodogram of the reduced and cleaned data set. Due to the irregularities of the thruster corrections, there remain some artefacts in the periodogram which we have not removed. The highest amplitude example of one of these artefacts occurs at about 84.5\,\cd\, with an amplitude of just 0.041\,mmag. The presence of these low-amplitude peaks does not affect our analysis.

\begin{figure}
  \includegraphics[width=\columnwidth]{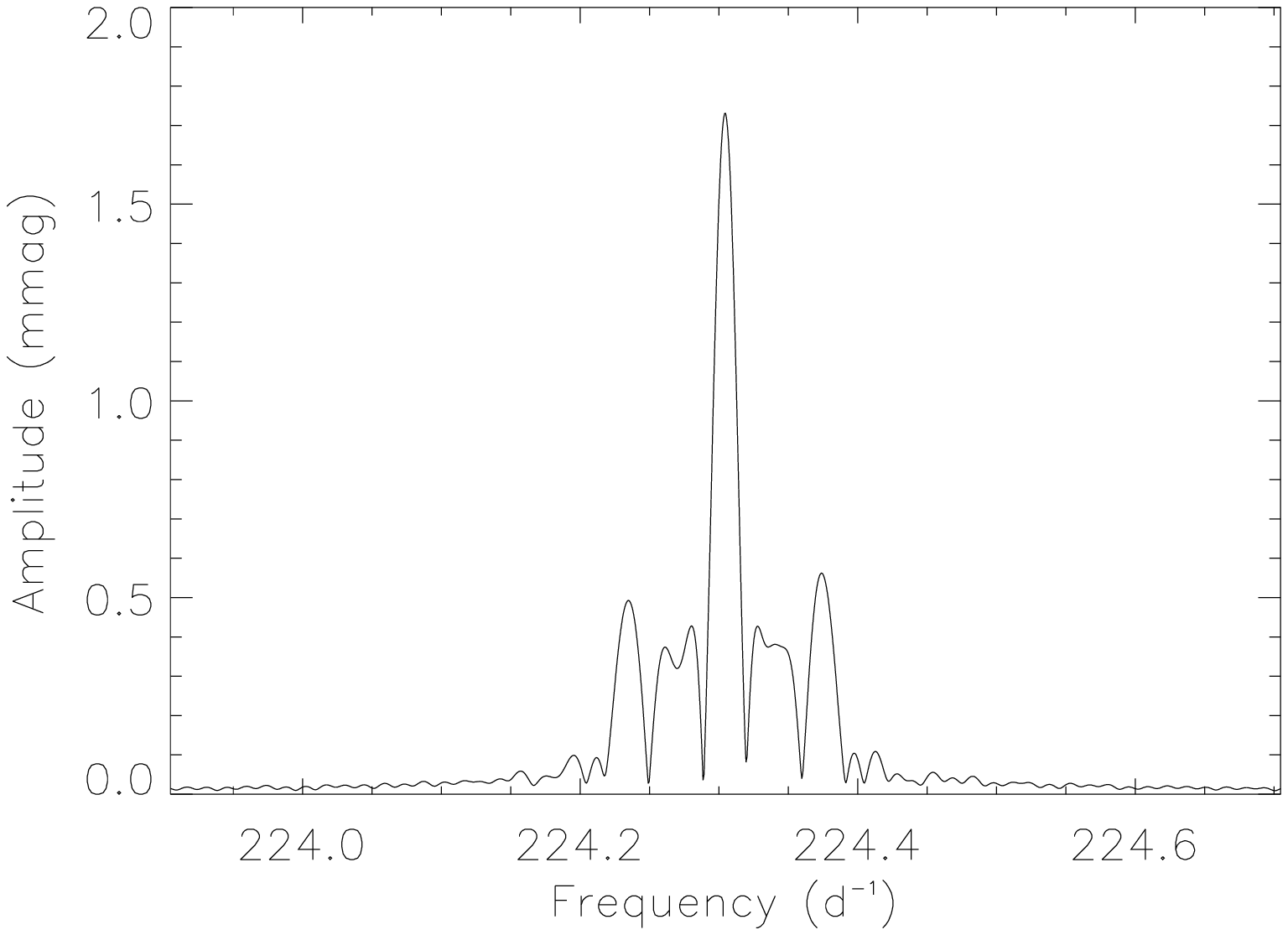}
  \includegraphics[width=\columnwidth]{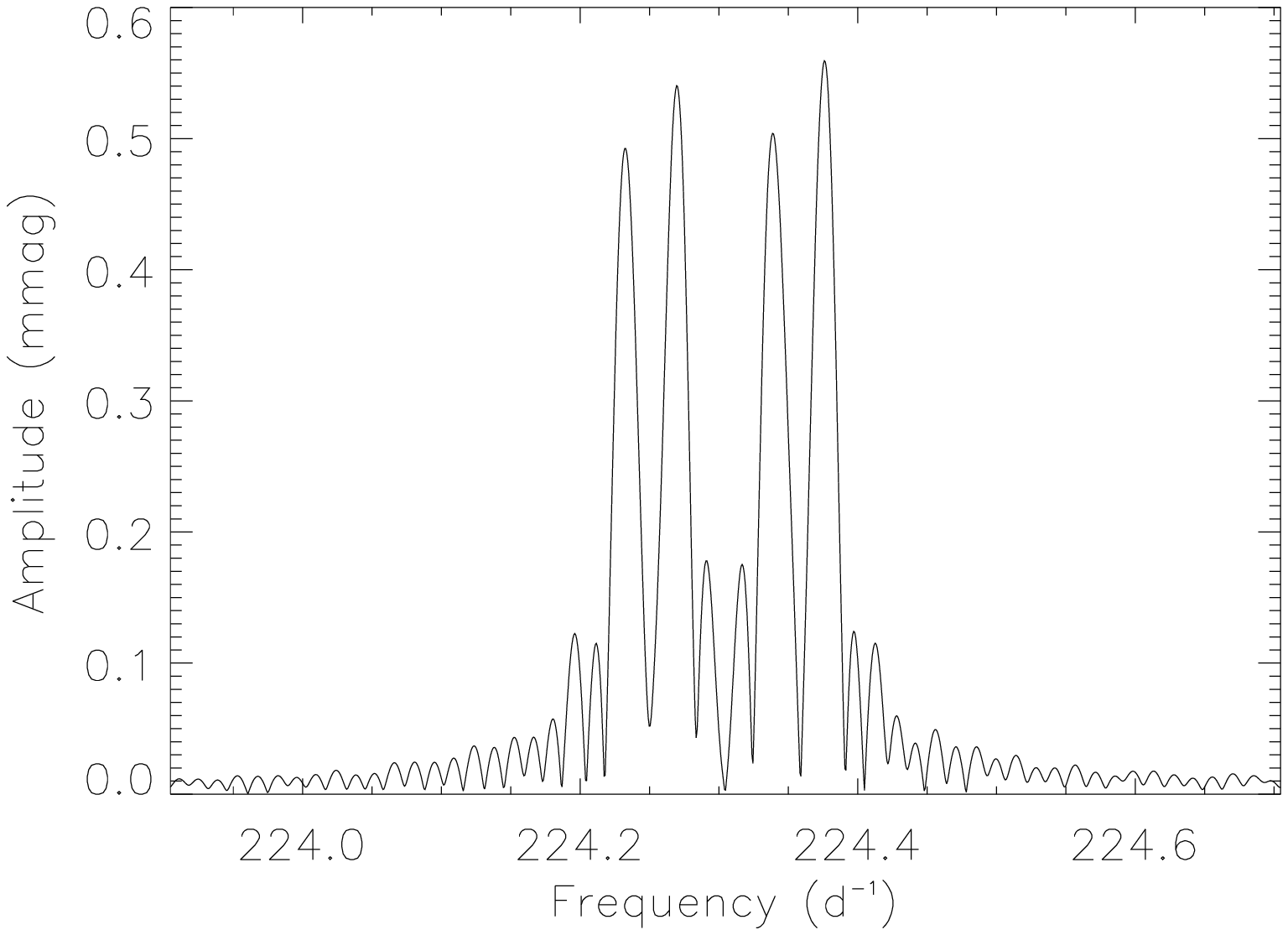}
  \caption{Top: the pulsation signature in the K2 SC data which shows a quintuplet separated by the rotation frequency. Bottom: the rotational sidelobes after the central peak, which is the actual pulsation frequency, has been removed. Analysis of the sidelobe amplitudes informs us on the geometry of the star (see text). Note the amplitude change between the two panels.}
  \label{fig:SC_pulse}
\end{figure}

\begin{figure*}
  \centering
  \includegraphics[width=\textwidth]{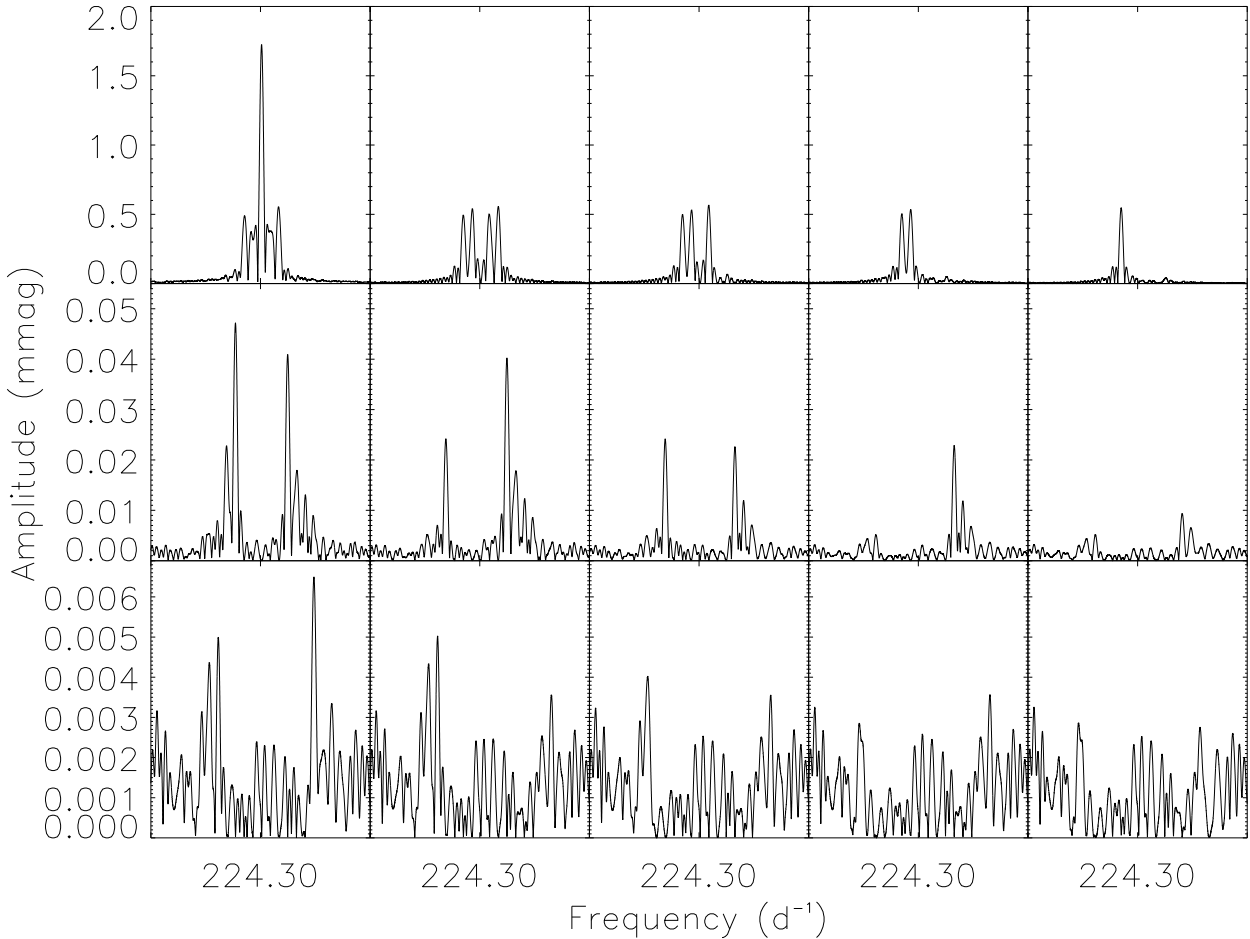}
  \caption{The pre-whitening of the pulsation and its sidelobes in the K2 data. The panels are read left to right, top to bottom. Note the change in amplitude scale on the different rows.}
  \label{fig:PW_K2}
\end{figure*}

Fig.\,\ref{fig:SC_pulse} shows the pulsation at a frequency of 224.3043\,\cd\, at an amplitude of 1.74\,mmag, and four of the sidelobes of the principal peak after it has been removed. The pulsation is similar to that seen in the WASP data (Fig.\,\ref{fig:lobes}), where there is a quintuplet split by the rotation period. Removing the four peaks shown in Fig.\,\ref{fig:SC_pulse} reveals further power at frequencies split by ever increasing multiples of the rotation frequency. In total, we identify 13 rotational sidelobes ($\nu+7\nu_{\rm rot}$ is below the noise level), the most ever seen in a roAp star. Fig.\,\ref{fig:PW_K2} demonstrates the pre-whitening of the sidelobes in the K2 data.

The results of a non-linear least-squares fit to the data are shown in Table\,\ref{tab:kplr_nlls}. The frequency difference, calculated in the final column of that table shows how closely the frequency splitting matches the rotation frequency of the star, as derived from the spots. Within the calculated errors, all sidelobes are in agreement (disregarding the peak at $\nu+8\nu_{\rm rot}$ due to the lack of a peak at $\nu+7\nu_{\rm rot}$). This shows that the oblique pulsation model is a good match to the observations of HD\,24355.

\begin{table*}
  \caption{The results of a non-linear least-squares fit to the K2 light curve. The last column shows the difference in the frequency between that line and the previous. The zero-point for the phases is BJD-2457097.2393.}
  \label{tab:kplr_nlls}
  \begin{tabular}{lccrc}
    \hline
    ID & Frequency & Amplitude & \multicolumn{1}{c}{Phase} & Frequency difference\\
       & (\cd)     & (mmag)    & \multicolumn{1}{c}{(rad)} & (\cd) \\
    \hline
$\nu-6\nu_{\rm rot}$ &$224.08925\pm0.00349$ & $0.004\pm0.002$ & $0.346 \pm0.444$ & \\		   
$\nu-5\nu_{\rm rot}$ &$224.12696\pm0.00289$ & $0.005\pm0.002$ & $-0.828\pm0.367$ & $0.03771\pm0.00453$\\
$\nu-4\nu_{\rm rot}$ &$224.16098\pm0.00059$ & $0.025\pm0.002$ & $-2.237\pm0.075$ & $0.03402\pm0.00295$\\
$\nu-3\nu_{\rm rot}$ &$224.19713\pm0.00032$ & $0.049\pm0.002$ & $2.051 \pm0.038$ & $0.03615\pm0.00067$\\
$\nu-2\nu_{\rm rot}$ &$224.23265\pm0.00003$ & $0.553\pm0.002$ & $2.116 \pm0.003$ & $0.03552\pm0.00032$\\
$\nu-1\nu_{\rm rot}$ &$224.26846\pm0.00003$ & $0.596\pm0.002$ & $-0.299\pm0.003$ & $0.03581\pm0.00004$\\
$\nu$              &$224.30430\pm0.00001$ & $1.863\pm0.002$ & $3.080 \pm0.001$ & $0.03584\pm0.00003$\\ 
$\nu+1\nu_{\rm rot}$ &$224.34011\pm0.00003$ & $0.599\pm0.002$ & $0.741 \pm0.003$ & $0.03581\pm0.00003$\\
$\nu+2\nu_{\rm rot}$ &$224.37597\pm0.00003$ & $0.627\pm0.002$ & $-2.264\pm0.003$ & $0.03586\pm0.00004$\\
$\nu+3\nu_{\rm rot}$ &$224.41145\pm0.00037$ & $0.044\pm0.002$ & $-1.599\pm0.042$ & $0.03548\pm0.00037$\\
$\nu+4\nu_{\rm rot}$ &$224.44717\pm0.00071$ & $0.023\pm0.002$ & $1.806 \pm0.082$ & $0.03572\pm0.00080$\\
$\nu+5\nu_{\rm rot}$ &$224.48330\pm0.00154$ & $0.010\pm0.002$ & $2.562 \pm0.191$ & $0.03613\pm0.00170$\\
$\nu+6\nu_{\rm rot}$ &$224.51895\pm0.00211$ & $0.007\pm0.002$ & $0.815 \pm0.268$ & $0.03565\pm0.00261$\\
$\nu+8\nu_{\rm rot}$ &$224.59382\pm0.00395$ & $0.004\pm0.002$ & $0.833 \pm0.499$ & $0.07487\pm0.00448$\\

\hline
  \end{tabular}
\end{table*}

The oblique pulsation model expects that the frequencies of the sidelobes are exactly split by the rotation frequency. Therefore, using the rotation period derived from the spot variations, we force the frequencies to be equally split, and perform a linear least-squares fit to the data. When performing this test, we set the zero-point in time such that the phases of the first sidelobes are equal. The results of this test are shown in Table\,\ref{tab:kplr_lls}. When the exact splitting of the sidelobes is set, we see that the phases of the central five peaks are similar. For a pure quadrupole mode, we expect a quintuplet of frequencies with all phases equal at the time of pulsation maximum. Since this is not the case here for the phases of the quintuplet frequencies, and because there are further rotational sidelobes, we conclude that the mode is distorted.

\begin{table}
  \caption{The results of a linear least-squares fit to the K2 light curve. The zero-point has been chosen to be BJD\,2457108.886701 to force the phases of the first rotational sidelobes to be equal. }
 \centering
  \label{tab:kplr_lls}
  \begin{tabular}{lccr}
    \hline
    ID & Frequency & Amplitude & \multicolumn{1}{c}{Phase} \\
       & (\cd)     & (mmag)    & \multicolumn{1}{c}{(rad)} \\
    \hline
$\nu-6\nu_{\rm rot}$ & 224.089368 & $0.004\pm0.002$ & $0.782 \pm0.445$ \\
$\nu-5\nu_{\rm rot}$ & 224.125190 & $0.005\pm0.002$ & $2.180 \pm0.367$ \\
$\nu-4\nu_{\rm rot}$ & 224.161012 & $0.025\pm0.002$ & $-2.850\pm0.074$ \\
$\nu-3\nu_{\rm rot}$ & 224.196834 & $0.049\pm0.002$ & $-2.225\pm0.037$ \\
$\nu-2\nu_{\rm rot}$ & 224.232656 & $0.553\pm0.002$ & $0.465 \pm0.003$ \\
$\nu-1\nu_{\rm rot}$ & 224.268478 & $0.596\pm0.002$ & $0.672 \pm0.003$ \\
$\nu$              & 224.304300 & $1.863\pm0.002$ & $0.388 \pm0.001$ \\
$\nu+1\nu_{\rm rot}$ & 224.340122 & $0.598\pm0.002$ & $0.671 \pm0.003$ \\
$\nu+2\nu_{\rm rot}$ & 224.375944 & $0.627\pm0.002$ & $0.288 \pm0.003$ \\
$\nu+3\nu_{\rm rot}$ & 224.411766 & $0.044\pm0.002$ & $-2.710\pm0.042$ \\
$\nu+4\nu_{\rm rot}$ & 224.447588 & $0.023\pm0.002$ & $-2.966\pm0.081$ \\
$\nu+5\nu_{\rm rot}$ & 224.483410 & $0.010\pm0.002$ & $0.397 \pm0.188$ \\
$\nu+6\nu_{\rm rot}$ & 224.519232 & $0.007\pm0.002$ & $1.267 \pm0.272$ \\
        \hline
  \end{tabular}
\end{table}

\subsection{Frequency variability and a rotational view of the pulsation geometry}
\label{sec:freq_var}

Many of the roAp stars that have been observed at such a high photometric precision as that afforded by the {\it{Kepler}} spacecraft have shown significant variations in their pulsation frequencies over the observation period \cite[e.g.][]{holdsworth14b,smalley15}. To this end, we investigate this phenomenon in HD\,24355. We split the data into sections of 100 pulsation cycles, or about 0.45\,d. This provides us with a suitable number of data points per section with which to calculate a periodogram over, and ample time resolution to detect any frequency variability. For this exercise, we use the data that have had the rotation signature, and its harmonics, removed (Fig.\,\ref{fig:K2_lc}, bottom).

\begin{figure*}
  \includegraphics[width=\textwidth]{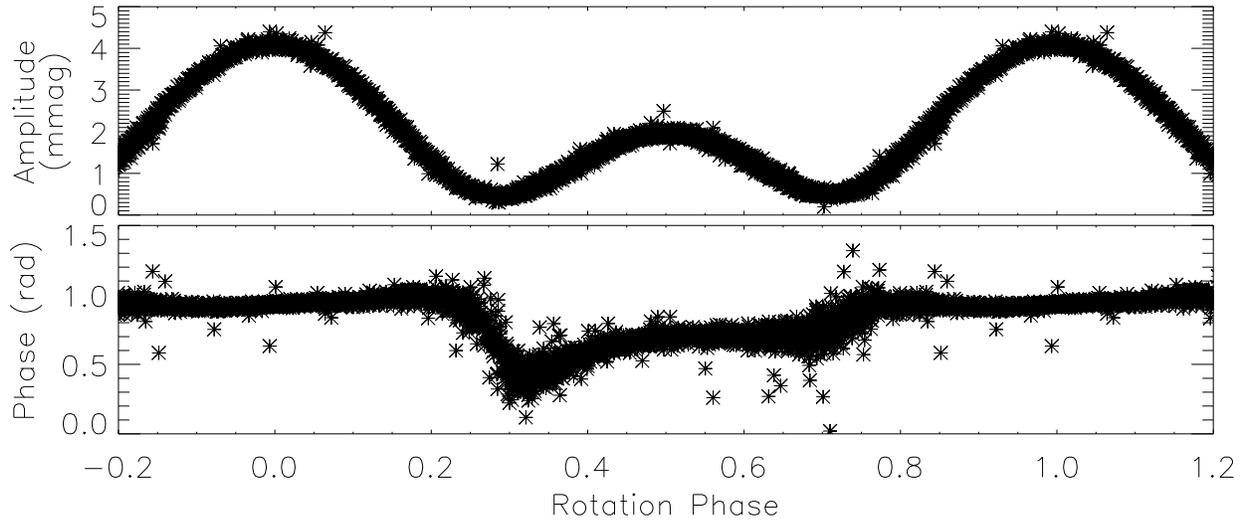}
  \caption{Top: the variation of the pulsation amplitude as a function of rotation phase. This shows that the pulsation amplitude coincides with the rotational light extremum. The double wave nature of the light curve is indicative of a pulsation mode seen from varying aspect where the line-of-sight crosses one pulsation node. In this case, the mode is a quadrupole. Bottom: the pulsation phase variation as a function of rotation phase. There is a phase change as the line-of-sight passes over pulsation nodes, however such a small change is indicative of an extremely distorted mode. The rotation phases are calculated as in equation\,(\ref{equ:rot}).}
  \label{fig:pulse_amp_phase}
\end{figure*}

By calculating the frequencies, amplitudes and phases of the pulsation in these sections of data, we are able to monitor the variations of the pulsation over the rotation period of the star. Fig.\,\ref{fig:pulse_amp_phase} shows the variation of the pulsation amplitude over the rotation period of the star. As is expected from the oblique pulsator model, the amplitude varies over the rotation cycle, as the pulsation mode is seen from varying aspects. The maximum amplitude coincides with light maximum which is expected when the spots producing the light variations are closely aligned with the magnetic and pulsation poles. The amplitude does not reach zero. If the mode were a pure mode, the amplitude would go to zero whenever the line-of-sight passes over a pulsation node; as this is not the case here, we can see that the mode is distorted. Other than the expected rotational amplitude variation, the pulsational amplitude is stable over the $\sim2.5$ rotations that the observations cover.

\begin{figure*}
  \centering
  \includegraphics[width=\textwidth]{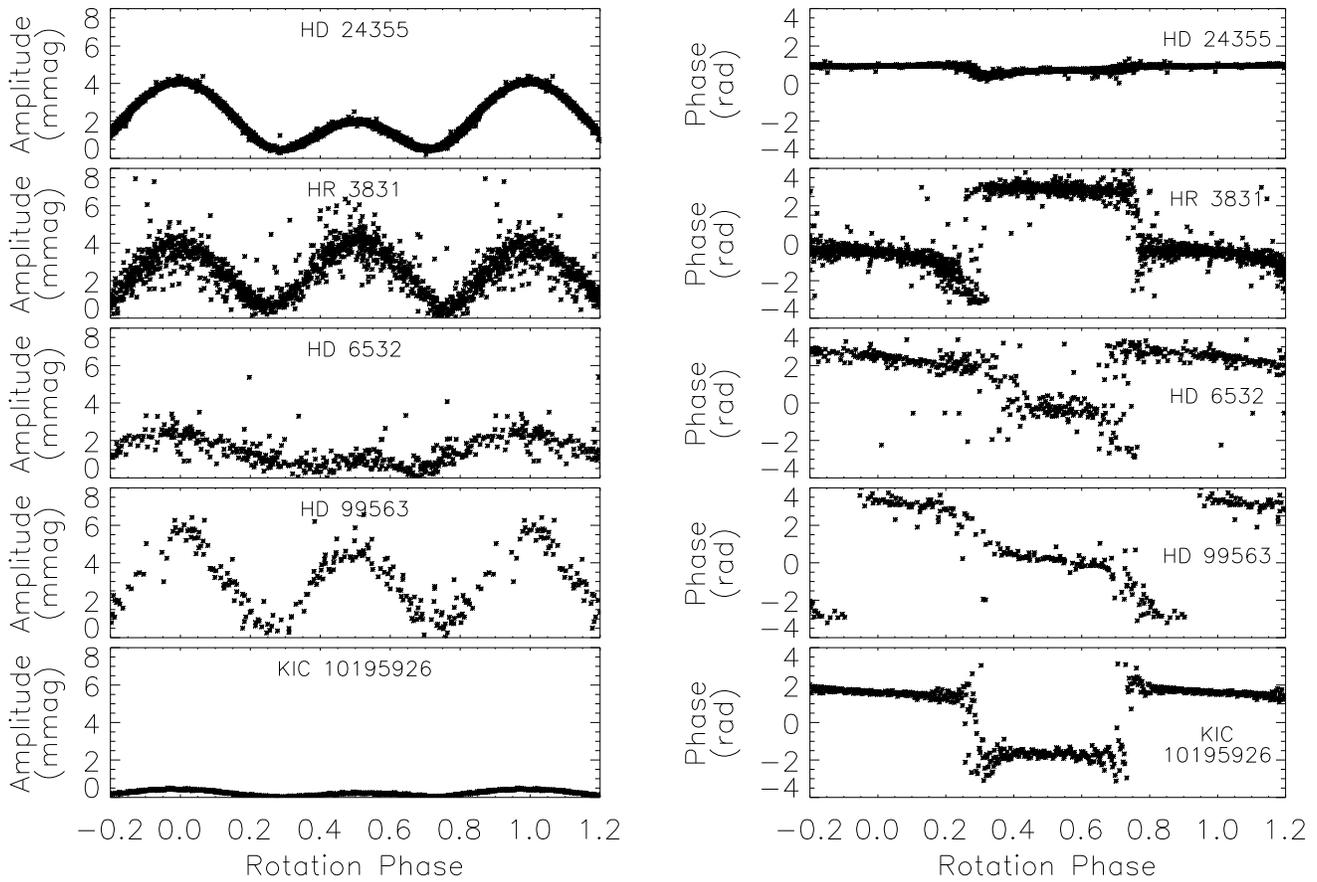}
  \caption{Pulsation amplitude and phase variations as a function of rotation phase for a set of roAp stars which are distorted dipole pulsators. All plots are on the same scale for easy comparison, and have been calculated in the same way. The stars are labelled in the plots. Despite all being distorted dipoles, there is a clear difference between the previously studied stars and HD\,24355. Note, the phases are different from those previously published for the comparison stars as a different zero-point was chosen.}
  \label{fig:other_phases}
\end{figure*}

The phase variations seen in Fig.\,\ref{fig:pulse_amp_phase} are very different to those seen in other roAp stars (e.g., see figure 10 of \citealt{kurtz11}). When the line-of-sight passes over a node, the oblique pulsator model predicts a $\pi-$rad phase shift for a pure dipole or quadrupole mode. This is not the case here, thus supporting the conclusion that HD\,24355 is a distorted pulsator. 

To compare this star with others which show distorted dipole pulsations, we have performed the same analysis on HR\,3831, HD\,6532, HD\,99563 and KIC\,10195926 \citep{kurtz97,kurtz96,handler06,kurtz11}. All of these stars, as well as HD\,24355, show a double wave rotationally modulated light curve, and phase changes at minimum light. Fig.\,\ref{fig:other_phases} shows this comparison. While the other stars show a clear phase change of about $\pi$\,rad at quadrature, the phase variations seen in HD\,24355 are dramatically different. Such an unusually small change in phase, and the number of sidelobes identified in the pulsation, show that is star has the most distorted pulsation mode of any roAp star yet observed. 

Beyond the known pulsation in HD\,24355, no further modes were identified in the K2 data. 

\subsubsection{Mode geometry constraints}

As the K2 data have a much higher precision than ground-based data, we can derive parameters of the geometry of the pulsation mode using the ratios of the sidelobes following the methodology of \citet{kurtz90}. Initially, we test the phase relations between the pulsation mode and the first rotational sidelobes by forcing the sidelobes to have the same phase. We find that $\phi_{-1}=\phi_{+1}\neq\phi_0$, suggesting that the mode is distorted (Table\,\ref{tab:kplr_lls}). 

The clear frequency quintuplet strongly suggests that the pulsation is quadrupolar. We therefore examine the axisymmetric quadrupole case, where $l=2$ and $m=0$, applying  a relationship from \citet{kurtz90} for a non-distorted oblique quadrupole pulsator in the absence of limb-darkening and spots, i.e.
\begin{equation}
  \tan i\tan\beta=4\frac{A^{(2)}_{+2}+A^{(2)}_{-2}}{A^{(2)}_{+1}+A^{(2)}_{-1}},
  \label{equ:quad}
\end{equation}
where $i$ is the rotational inclination angle, $\beta$ is the angle of obliquity between the rotation and magnetic axes, and $A^{(2)}_{\pm1}$ and $A^{(2)}_{\pm2}$ are the amplitudes of the first and second sidelobes of the quadrupole pulsation, respectively. Using the amplitudes from Table\,\ref{tab:kplr_lls} again, and substituting them into equation\,(\ref{equ:quad}), we derive $\tan i\tan\beta=3.950\pm0.012$ for this simplified case.

Table\,\ref{tab:i_beta} shows a set of values of $i$ and $\beta$ that satisfy the constraint of equation\,(\ref{equ:quad}). All possible values sum to greater than 90$^\circ$, which is consistent with the rotational light modulation under the assumption that there are spots at both magnetic/pulsation poles. Values of $i$ and $\beta$ near to 90$^\circ$ are not permitted for a quadrupole mode, within the constraint of equation\,(\ref{equ:quad}), because the line-of-sight would then not pass over a node. However, we have a stronger constraint on the inclination: from our spectra we are able to place an upper limit of $v\sin i\leq3.5$\,\kms. Using this value, the rotation period of $27.9158$\,d and the derived radius of $R=2.53\,{\rm R_{\odot}}$, we can estimate an upper limit of $i \le 50^{\circ}$.

\begin{table}
    \caption{Values of $i$ and $\beta$ which satisfy equation\,(\ref{equ:quad}). We list values up to $i=\beta$, where the values of $i$ and $\beta$ are reversed. The columns $i-\beta$ and $i+\beta$ show the extreme values of the angle between the line-of-sight and the pulsation pole; these can be useful in visualising the pulsation geometry.}
    \centering
    \label{tab:i_beta}
  \begin{tabular}{cccc}
    \hline
    $i$ & $\beta$ & $i-\beta$ & $i+\beta$ \\
    \hline
    89.0 & 3.9 & 85.1 & 92.9 \\
    85.0 & 19.0 & 66.0 & 104.0 \\
    80.0 & 34.9 & 45.1 & 114.9 \\
    75.0 & 46.6 & 28.4 & 121.6 \\
    70.0 & 55.2 & 14.8 & 125.2 \\
    65.0 & 61.5 & 3.5  & 126.5 \\
    63.1 & 63.1 & 0.0  & 126.2 \\
    \hline
  \end{tabular}
\end{table}

Using these values for $i$ and $\beta$ we apply the method of \citet{kurtz92}, based on work by \citet{st93}, to deconvolve the pulsation into the components of a spherical harmonic series. This technique separates the distorted mode into its pure $\ell=0,1,2,\ldots$ spherical harmonic components, allowing us to see the shape of the mode. The results of this deconvolution are shown in Tables\,\ref{tab:har} and \ref{tab:har_comp}. The results show that the mode is a quadrupole mode with a strong spherically symmetric distortion represented by the radial component. The $\ell=1$ and $3$ components are small in comparison.

\begin{table}
  \caption{Components of the spherical harmonic series description of the pulsation for $i=45^\circ$ and $\beta=77^\circ$.}
  \centering
  \label{tab:har}
  \begin{tabular}{lcccc}
    \hline
    $\ell$ & 0 & 1 & 2 & 3 \\
    \hline
    
    $A_{-3}^{(\ell)}$ (mmag) &        &        &        & 0.049 \\
    $A_{-2}^{(\ell)}$ (mmag) &        &        & 0.623  & 0.066 \\
    $A_{-1}^{(\ell)}$ (mmag) &        & 0.123  & 0.629  & -0.067 \\
    $A_0^{(\ell)}$ (mmag)   &  2.353 & 0.063  & -0.416 & 0.025 \\ 
    $A_{+1}^{(\ell)}$ (mmag) &        & 0.148  & 0.661  & -0.064 \\
    $A_{+2}^{(\ell)}$ (mmag) &        &        & 0.678  & 0.062 \\
    $A_{+3}^{(\ell)}$ (mmag) &        &        &        & 0.044 \\
    $\phi$  (rad)        & 0.007  & 2.664  & 0.114  & -2.621 \\

    \hline

  \end{tabular}
\end{table}

\begin{table}
  \caption{Comparison between the observed amplitudes and phases to those calculated with the spherical harmonic fit.}
  \centering
  \label{tab:har_comp}
  \begin{tabular}{lccrr}
    \hline
    ID & $A_{\rm obs}$ & $A_{\rm calc}$ & \multicolumn{1}{c}{$\phi_{\rm obs}$} & \multicolumn{1}{c}{$\phi_{\rm calc}$} \\
    \hline
    $\nu-3\nu_{\rm rot}$ & $0.049\pm0.002$ & 0.049 & $ -2.621\pm0.038$ & -2.621 \\
    $\nu-2\nu_{\rm rot}$ & $0.553\pm0.002$ & 0.553 & $  0.067\pm0.003$ & 0.067 \\
    $\nu-1\nu_{\rm rot}$ & $0.596\pm0.002$ & 0.596 & $  0.275\pm0.003$ & 0.275 \\
    $\nu$              & $1.863\pm0.002$ & 1.863 & $ -0.008\pm0.001$ & -0.008 \\ 
    $\nu+1\nu_{\rm rot}$ & $0.598\pm0.002$ & 0.603 & $  0.275\pm0.003$ & 0.288 \\
    $\nu+2\nu_{\rm rot}$ & $0.627\pm0.002$ & 0.621 & $ -0.108\pm0.003$ & -0.076 \\
    $\nu+3\nu_{\rm rot}$ & $0.044\pm0.002$ & 0.044 & $ -3.107\pm0.042$ & -2.621 \\
    \hline
  \end{tabular}
\end{table}

Similarly to the APT data, the K2 data show harmonics of the pulsation. These harmonics are present up to and including the 4$^{\rm th}$, i.e. $5\nu = 1121.5215$\,d$^{-1}$, where at higher frequencies any signal is lost in the noise. This demonstrates the non-sinusoidal nature of pulsations in roAp stars.

The opportunity to observe a high-amplitude, classically ground-based, roAp star with K2 precision provided, for the first time, the chance to compare the space- and ground-based targets. It is unclear, due to the detection limits from the ground, and the small sample of stars observed at $\muup$mag precision, whether there is a fundamental difference between the high-amplitude ground-based targets and the low-amplitude space-based targets. The observations presented here, although for only one of the 57 ground-based targets, suggest that there may be an `amplitude desert' in the roAp stars. However, many more observations of ground-based roAp stars at $\muup$mag precision are needed to test whether there really is such a dichotomy. 

\subsubsection{Comparison of K2 and APT data}
\label{sec:APT-K2}

The availability of both the broadband K2 observations and the three nights of $B-$band APT data allows us to directly compare the amplitudes of the different data sets, and estimate the maximum pulsation amplitude of HD\,24355 in a $B-$band filter. From Table\,\ref{tab:apt}, we know the rotation phase at which the APT data were obtained, namely at phases 0.873, 0.907 and 0.943. We compare the amplitudes of the APT data at these phases, to the amplitudes of the K2 data at the same phases in Table\,\ref{tab:amp_comp}. We can see that the amplitude ratio is increasing with time, so rather than take an average of the ratio, we use a linear extrapolation to the maximum of the pulsation in the K2 data, which is measured to be $A_{\rm K2}=4.10\pm0.09$. This gives us a maximum $B-$band amplitude of $6.90\pm1.93$\,mmag. Comparing this amplitude with other roAp stars in the literature for which $B-$band amplitudes are known \citep[e.g.][]{kurtz06,holdsworth15}, this is not a surprising result for HD\,24355.

\begin{table}
\centering
  \caption{Comparison of the APT $B$-band observations from 2013 November and K2 pulsation amplitudes at given rotation phases.}
  \label{tab:amp_comp}
  \begin{tabular}{cccc}
    \hline
    Rotation & $A_{\rm APT}$ & $A_{\rm K2}$ & $A_{\rm APT}/A_{\rm K2}$ \\
    Phase & (mmag) & (mmag) & \\
    \hline
    0.873 & $3.52\pm0.17$ & $2.63\pm0.13$ & $1.34\pm0.09$\\
    0.907 & $4.79\pm0.15$ & $3.25\pm0.06$ & $1.47\pm0.05$\\
    0.943 & $5.84\pm0.16$ & $3.64\pm0.05$ & $1.60\pm0.05$\\
    \hline
  \end{tabular}
\end{table}

Further to the three nights of observations in 2013 November, we observed HD\,24355 with the APT simultaneously with K2. These coincident observations were made in three filters, $U$, $B$ and $V$. Observations we taken as in 2013, but cycling through each of the three filters sequentially. We attempted to observe the star throughout the K2 observations, however due to the poor visibility of the target, from Earth, during the K2 campaign we were unable to obtain long periods on target, recording only 665 data points per filter over a 26-d period. These multi-colour data are still useful. We combine the two best nights of data, namely BJD $245\,7091.6$ and $245\,7092.6$, and fit the pulsation by non-linear least-squares. The results of the fitting are presented in Table\,\ref{tab:APT-col}. To compare the multi-colour data to K2 observations, we extract the same time period from the K2 data and compute the pulsation amplitude in the same way. Again, it is clear that the amplitude seen in the white light {\textit{Kepler}} filter is greatly reduced when considering the $B$, or even $U$, band observations, an expected result from multi-colour photometric studies \citep{medupe98}. 

We observe here that the $B$-band amplitude at $\phi_{\rm rot}=0.400$ is greater than the K2 amplitude by a factor of $4.34\pm0.47$. This is a much greater ratio than the observations at $\phi_{\rm rot}=0.943$, close to maximum pulsation amplitude. For the values of $i$ and $\beta$ which produce the best model to the data (see Section\,\ref{sec:model}), namely $i=45^\circ$ and $\beta=77^\circ$, we see the pulsation poles at $32^\circ$ and $58^\circ$ to the line-of-sight, corresponding to the pulsation primary and secondary maxima, respectively. We postulate, therefore, that the variation in the amplitude ratio between $B$-band and K2 observations is a result of a different line-of-sight to the pulsation pole, and as such we are probing higher in the atmosphere where pulsation amplitudes are greater for certain spectral lines \cite[e.g.][]{kurtz06b}, and possibly in broad-band photometry too. We would then expect the $B$-band amplitude of the secondary light maximum at $\phi_{\rm rot}=0.5$ to be of the order $8.65\pm0.94$\,mmag, thus shifting the maximum amplitude, in the $B$-band, by half a rotation period. To test the relation between pulsation amplitude, rotation phase (or light-of-sight aspect) and atmospheric depth requires multi-colour photometry covering the entire rotation period of the star.

\begin{table}
\centering
\caption{APT multi-colour photometry of HD\,24355, taken at rotation phase 0.400.}
\label{tab:APT-col}
\begin{tabular}{ccc}
\hline
Filter & Amplitude & $A_{\rm APT}/A_{\rm K2}$\\
 & (mmag) & \\
\hline
$U$ & $3.02\pm0.78$ & $2.54\pm0.66$\\
$B$ & $5.17\pm0.51$ & $4.34\pm0.47$\\
$V$ & $1.67\pm0.48$ & $1.40\pm0.41$\\
K2 & $1.19\pm0.05$ & \\
\hline
\end{tabular}
\end{table}

\section{Modelling the amplitude and phase modulation}
\label{sec:model}

\subsection{Equations}
We assume that the rapid pulsation of HD\,24355 is an axisymmetric (with respect to the magnetic axis) high-order p mode. The presence of four main rotational sidelobes frequencies indicates the pulsation to be a quadrupole mode. However, in the presence of a magnetic field (assuming a dipole field) no pure quadrupole mode exists, because the Legendre function $P_\ell(\cos\theta_B)$ couples to another $P_{\ell'}(\cos\theta_B)$ if $\ell'=\ell\pm 2$; i.e., the magnetic field distorts the angular distribution of amplitude.

The distribution of the local luminosity perturbation on the surface may be expressed by a sum of terms proportional to $Y_{\ell_j}^0(\theta_B,\phi_B)  [ =N_{\ell_j 0}P_{\ell_j}(\cos\theta_B)]$ with even degrees $\ell_j = 0, 2, 4, \ldots$ as
\begin{equation}
\delta F(\theta_{\rm B},\phi_{\rm B},t)
={\rm e}^{{\rm i}\sigma t}\sum_{j=1}^J f_j N_{\ell_j0}P_{\ell_j}(\cos\theta_{\rm B}) ,
\label{eq:delF}
\end{equation}
where $\ell_j = 2(j-1)$, $N_{\ell_j0}=\sqrt{(2\ell_j+1)/(4\pi)}$, and $(\theta_{\rm B},\phi_{\rm B})$ are spherical coordinates whose axis is along the magnetic axis. The luminosity perturbation at the surface for each component $f_j$ is obtained by solving non-adiabatic non-radial pulsation equations (truncated at $j=J$) which include the effect of a dipole magnetic field \citep{saio05}. In the absence of magnetic fields, $f_j =0$ if $j\not=2$ for quadrupole modes, while magnetic couplings make $f_j\not=0$ for any $j$. 

Observational light variation may be expressed as
 \begin{equation}
  \begin{array}{l} \displaystyle
  \Delta L(t)  \propto
  {\rm e}^{{\rm i}\sigma t}\sum_{j=1}^J f_j N_{\ell_j,0} \int_0^{2\pi} d\phi_{\rm L} 
  \cr \displaystyle \qquad
  \times \int_0^{\pi/2}\!\!\! d\theta_{\rm L}\sin\theta_{\rm L}(1-\mu+\mu\cos\theta_{\rm L})\cos\theta_{\rm L}
    P_{\ell_j}(\cos\theta_{\rm B}),
  \end{array}
  \label{eq:delL}
 \end{equation}
 where $\mu$ is the limb-darkening parameter, and $(\theta_{\rm L},\phi_{\rm L})$ is the set of spherical coordinates on the surface, whose axis ($\theta_{\rm L} = 0$) is along the observer's line-of-sight. Numerical integration over the visible hemisphere can be done by using the relation between $(\theta_{\rm L},\phi_{\rm L})$ and $\cos\theta_{\rm B}$, which is obtained from spherical trigonometry \citep[e.g.,][]{smart31}. The amplitude modulation occurs because the relation is time-dependent (the coordinate $(\theta_{\rm L},\phi_{\rm L})$ at a magnetic pole changes with rotational phase).
 
We have neglected to consider the departure from spherical symmetry, which results from the presence of a strong magnetic field, in the determination of the luminosity variation. This effect can be safely disregarded as the ratio between the temperature variation and the radius variation is of the order 100 for high order p-modes, thus temperature variations are much more significant than radius variations caused by the strong magnetic field.

Another way to evaluate equation\,(\ref{eq:delL}) is to use relations of spherical harmonics associated with polar coordinates with different axes \citep[see][]{st93,saio04, unno89}. The spherical harmonic $Y_{\ell_j}^0(\theta_{\rm B},\phi_{\rm B})$ is converted to a sum of terms proportional to $Y_{\ell_j}^{m''}(\theta_{\rm L},\phi_{\rm L})$ as
\begin{equation}
\begin{array}{ll}\displaystyle
Y_{\ell_j}^0(\theta_{\rm B},\phi_{\rm B}) 
& \displaystyle
= \sum_{m'=-\ell_j}^{\ell_j}Y_{\ell_j}^{m'}(\theta_{\rm R},\phi_{\rm R})d^{(\ell_j)}_{m',0}(\beta)
\cr & \displaystyle
=\sum_{m'=-\ell_j}^{\ell_j}Y_{\ell_j}^{m'}(\theta_{\rm I},\phi_{\rm I})d^{(\ell_j)}_{m',0}(\beta)
{\rm e}^{-{\rm i}m'\Omega t}
\cr & \displaystyle 
=\sum_{m'=-\ell_j}^{\ell_j}d^{(\ell_j)}_{m',0}(\beta)
{\rm e}^{-{\rm i}m'\Omega t} \!\!\! \sum_{m''=-\ell_j}^{\ell_j} \!\! Y_{\ell_j}^{m''}(\theta_{\rm L},\phi_{\rm L})d^{(\ell_j)}_{m'',m'}(i),
\end{array}
\label{eq:ylm}
\end{equation}
where coordinate $(\theta_{\rm R},\phi_{\rm R})$ is the co-rotating frame with the axis along the rotation axis, which is inclined to the magnetic axis by angle $\beta$. While the axis of the system $(\theta_{\rm I},\phi_{\rm I})$ is also the rotation axis, this is an inertial system; i.e., $\theta_{\rm I} = \theta_{\rm R}$ and $\phi_{\rm I}= \phi_{\rm R} + \Omega t$. The last equality of equation\,(\ref{eq:ylm}) relates the system $(\theta_{\rm I},\phi_{\rm I})$ to the system $(\theta_{\rm L},\phi_{\rm L})$ whose axis is along the line-of-sight and inclined to the rotation axis by angle $i$. Substituting the last relation in equation\,(\ref{eq:ylm}), we obtain
\begin{equation}
\Delta L(t)  \propto
{\rm e}^{{\rm i}\sigma t}\sum_{j=1}^J f_j N_{\ell_j0} I_{\mu\ell_j} \!\!\! \sum_{m'=-\ell_j}^{\ell_j} \!\!\!d^{(\ell_j)}_{m',0}(\beta)d^{(\ell_j)}_{0,m'}(i){\rm e}^{-{\rm i}m'\Omega t},
\label{eq:sidelobes}
\end{equation}
where $I_{\mu\ell_j}$ represents the latitudinal integration
\begin{equation}
I_{\mu\ell_j} \equiv  \int_0^1 (1-\mu+\mu z)zP_{\ell_j}(z)dz .
\end{equation}
In this expression, amplitude modulation is expressed by a superposition of Fourier components at frequencies of $\sigma \pm k\Omega$ with $k = 0, 1, \ldots$, corresponding to the observational data given in Table\,\ref{tab:kplr_lls}.

\subsection{Comparison with observation}
Spectroscopic parameters of HD\,24355  are $(T_{\rm eff}, \log g) = (8200 \pm 200\,{\rm K}, 4.0 \pm 0.2)$.  Within and around the error-box we have searched for models that have a frequency similar to 224.30\,d$^{-1}$ and reproduce the amplitude and phase modulation.

\begin{figure}
  \centering
\includegraphics[width=\columnwidth]{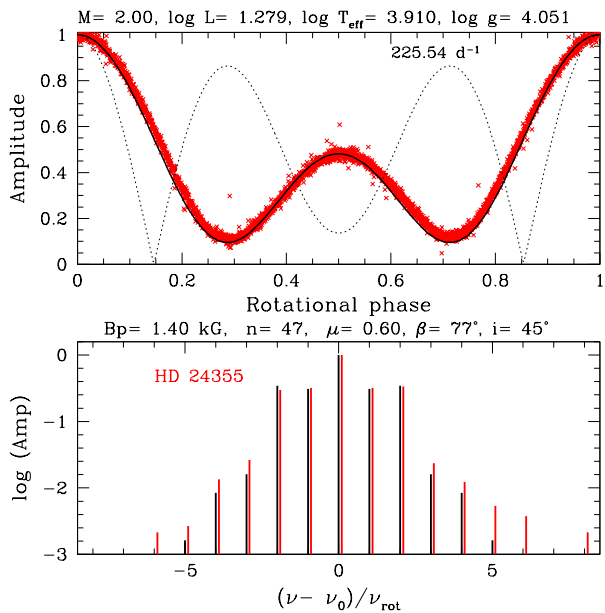}
\caption{Amplitude modulation (top panel) and corresponding rotational sidelobes (bottom panel) of HD\,24355 (red crosses and red bars) are compared with a model (black lines). The dotted line in the top panel shows amplitude modulation expected if the latitudinal distribution of pulsation were proportional to the Legendre function $P_2(\cos\theta_B)$.
Note that bottom panel plots the logarithmic (base 10) values of sidelobes amplitude normalised by the amplitude of the central frequency. 
}
\label{fig:ampmod_model}
\end{figure}

For each model we picked a non-magnetic quadrupole mode having a frequency similar to the observed frequency $224.3$\,d$^{-1}$. Then, assuming a polar magnetic field strength, $B_{\rm p}$ of a dipole field, we obtained non-adiabatic eigenfrequency and eigenfunction using the method discussed in \citet{saio05}, in which the expansion was truncated at $J= 20$ (equation\,\ref{eq:delF}). 
Using the eigenfunction in equation\,(\ref{eq:sidelobes}), we obtain the amplitude modulation and the amplitudes of rotational sidelobes to compare with the observations. A good fit, as presented in Fig.\,\ref{fig:ampmod_model}, was sought by changing $B_{\rm p}$,  $\beta$, and $i$ assuming the limb-darkening parameter $\mu = 0.6$ (the effect of different $\mu$ was examined by adopting $\mu =0.4$ and $0.8$ and found that qualitative properties hardly changed). Theoretical prediction depends weakly on the inclination $i$. This is constrained by the fact that the rotation period of HD\,24355 is 27.9158\,d and an upper limit of  $v\sin i \le 3.5$\,km\,s$^{-1}$.  As our spectroscopic observations allow us to place an upper limit of $i < 50^\circ$, we present models with $i = 45^\circ$ as a standard choice, although we can obtain similarly good models in a wide range of $i$.

\begin{figure}
  \centering
  \includegraphics[width=\columnwidth]{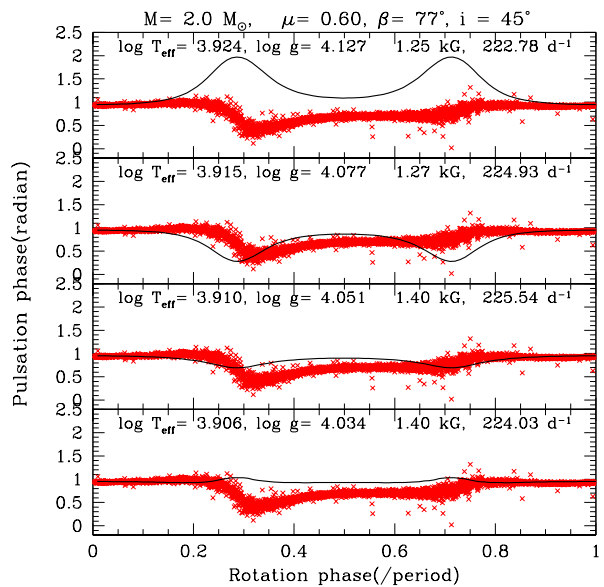}
\caption{Pulsation phase modulation of HD\,24355 (red crosses) with respect to the rotation phase is compared with theoretical ones obtained for some of the evolutionary models of $2\,{\rm M}_\odot$. The effective temperature and surface gravity decrease from top to bottom. 
}
\label{fig:phasemod_model}
\end{figure}

Although the amplitude modulation of HD\,24355 can be reasonably reproduced by models with parameters around the spectroscopically derived $(T_{\rm eff}, \log g)$, qualitative properties of the phase modulation systematically change along evolutionary tracks (i.e., $T_{\rm eff}$), as shown in Fig.\,\ref{fig:phasemod_model}; there is a point where the direction of theoretical phase change switches; the ``amplitude" of phase modulation decreases as $T_{\rm eff}$ decreases.  In producing Fig.\,\ref{fig:phasemod_model} we ensure that every model in each panel of the figure reproduces well the observed amplitude modulation, as in Fig.\,\ref{fig:ampmod_model}. We judge that the model in the second panel from the bottom at $\log T_{\rm eff}=3.910$ best fits with the observed phase modulation among the $2.0\,{\rm M}_\odot$ models. This model has a radius of $2.21\,{\rm R}_\odot$, which gives an upper limit of $i \le 48^\circ$ consistent with $i=45^\circ$.

\begin{figure}
\centering
\includegraphics[width=\columnwidth]{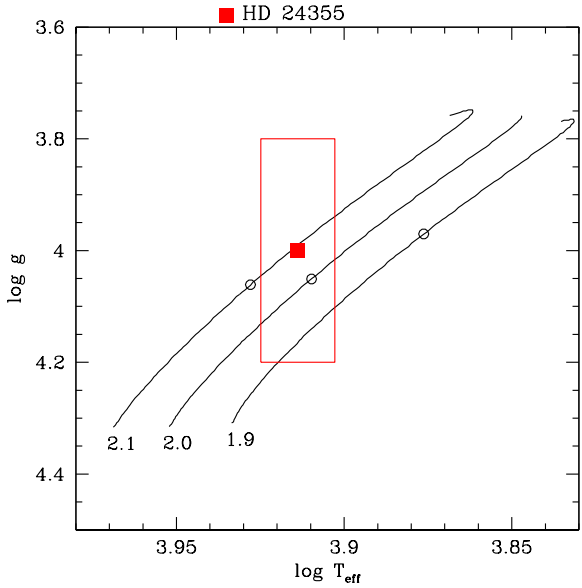}
\caption{The $\log T_{\rm eff} - \log g$ diagram showing the spectroscopic parameter, $(T_{\rm eff},\log g) = (8200 \pm 200, 4.0\pm0.2)$ of HD\,24355 and evolutionary tracks of $1.9, 2.0,$ and $2.1\,{\rm M}_\odot$ with a composition of $(X,Z) = (0.70,0.02)$. The open circle along each track indicates the model that can reproduce the amplitude and the phase modulations of HD\,24355. }
\label{fig:telg}
\end{figure}

In this way, for each evolutionary track, we determine a model which reasonably reproduces the amplitude modulation and the phase modulation of HD\,24355. The parameters of those models are shown by open circles in Fig.\,\ref{fig:telg}. Among the models we examined, the 2.0-M$_\odot$ model is consistent with the spectroscopic parameters of HD\,24355. Furthermore, a relatively weak magnetic field of $B_{\rm p} = 1.4$\,kG is consistent with the absence of Zeeman splitting in the spectral lines. Such a moderate magnetic field can deform the eigenfunction significantly, because the oscillation frequency of HD\,24355 corresponds to a very high-order p mode whose energy is confined in superficial layers of the star.

We note that a slight asymmetry in observed phase modulation, which cannot be reproduced in our models, is probably caused by a small effect of Coriolis force \citep{bigot11} which we have neglected for simplicity.

\begin{figure}
\centering
\includegraphics[width=\columnwidth]{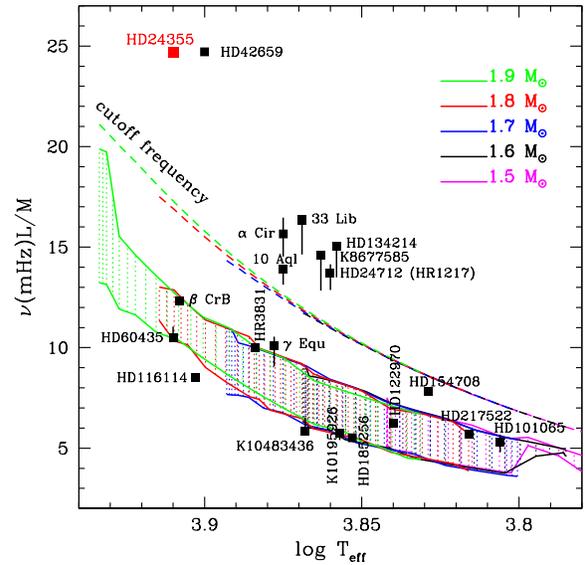}
\caption{The position of HD\,24355 in the $\log T_{\rm eff} - \nu L/M$ plane is compared with some of the known roAp stars, in which $L/M$ is in solar units. Main frequencies are shown by squares and frequency ranges by vertical bars.
Shaded region indicates where high-order p-modes are excited by the $\kappa$-mechanism in the H-ionisation zone in non-magnetic models. Dashed lines show the acoustic cut-off frequencies. This figure is an extended version of figure 3 in \citet{saio14}.
}
\label{fig:TefreqLM}
\end{figure}

Fig.\,\ref{fig:TefreqLM} shows the position of HD\,24355 in the $\log T_{\rm eff} - \nu L/M$ plane together with some of the known roAp stars (details of which can be found in Table\,\ref{tab:TefreqLM}). In this diagram, the mass dependence of the acoustic cut-off frequency as well as the range of excitation by the $\kappa$-mechanism in the H-ionisation zone \citep{balmforth01} are partially compensated for. This figure shows that HD\,24355 is a peculiar roAp star whose pulsation is highly super-critical -- much more so than for the stars in the well known cooler roAp group consisting of HR\,1217, $\alpha$\,Cir and 10\,Aql, etc., all of which are multi-periodic, while HD\,24355 is a single-mode pulsator. Another single mode roAp star, HD\,42659 \citep{martinez94}, could have similar properties to HD\,24355, although the position on the diagram depends on the uncertain quantity of $L/M$, and low-amplitude oscillations of HD\,42659 are not well studied. The excitation mechanism of such super-critical high-order p-modes is not known.

\begin{table*}
  \centering
  \caption{Parameters of the roAp stars used to produce Fig.\,\ref{fig:TefreqLM}.}
  \label{tab:TefreqLM}
  \begin{tabular}{lllllll}
    \hline
    Star name &  Main frequency &    Minimum frequency & Maximum frequency & Mass     & $\log(L/{\rm L_{\odot}})$  & $\log(T_{\rm eff})$ \\
    &        mhz      &      mhz             & mhz               & M$_\odot$ &         & K \\
    \hline
    HD\,24355 & 2.596 & 2.596 & 2.596 & 2.00 & 1.28 & 3.910 \\
HD\,24712 (HR\,1217)&  2.721     & 2.553     & 2.806    & 1.65  & 0.92   & 3.860 \\ 
HR\,3831         &  1.428     & 1.428     & 1.428    & 1.76  & 1.09   & 3.884 \\  
HD\,101065       &  1.373     & 1.246     & 1.474    & 1.53  & 0.77   & 3.806 \\ 
HD\,116114       &  0.782     & 0.782     & 0.782    & 1.92  & 1.32   & 3.903 \\
HD\,122970       &  1.501     & 1.501     & 1.501    & 1.48  & 0.79   & 3.840 \\ 
$\alpha$\,Cir   &  2.442     & 2.265     & 2.567    & 1.71  & 1.04   & 3.875 \\  
HD\,134214       &  2.950     & 2.647     & 2.983    & 1.65  & 0.93   & 3.858 \\   
$\beta$\,CrB    &  1.031     & 1.031     & 1.031    & 2.10  & 1.40   & 3.908 \\  
33\,Lib          &  2.015     & 1.803     & 2.015    & 1.78  & 1.16   & 3.869 \\  
10\,Aql          &  1.448     & 1.366     & 1.469    & 1.85  & 1.25   & 3.875 \\ 
$\gamma$\,Equ   &  1.365     & 1.228     & 1.427    & 1.70  & 1.10   & 3.878 \\ 
HD\,217522       &  1.200     & 1.200     & 1.200    & 1.49  & 0.85   & 3.816 \\  
K\,8677585     &  1.660     & 1.458     & 1.676    & 1.80  & 1.20   & 3.863 \\ 
HD\,42659        &  1.718     & 1.718     & 1.718    & 2.10  & 1.48   & 3.900 \\  
HD\,60435        &  1.381     & 1.381     & 1.457    & 1.82  & 1.14   & 3.910 \\  
HD\,154708       &  2.088     & 2.088     & 2.088    & 1.43  & 0.73   & 3.829 \\ 
HD\,185256       &  1.613     & 1.613     & 1.613    & 1.40  & 0.68   & 3.853 \\ 
K\,10483436      &  1.353     & 1.353     & 1.512    & 1.60  & 0.84   & 3.868 \\   
K\,10195926      &  0.975     & 0.920     & 0.975    & 1.70  & 1.00   & 3.857 \\ 
\hline
  \end{tabular}
\end{table*}

\section{Summary and Conclusions}

We have presented the first {\it Kepler} spacecraft observations of the high-amplitude roAp star HD\,24355 alongside a detailed analysis of the ground-based discovery and follow-up photometric data. The K2 data have allowed us to unambiguously determine the rotation period of the star to be $27.9158\pm0.0043$\,d, a parameter which was uncertain when considering the ground-based SuperWASP data alone. Classification dispersion spectra allowed us to classify this star as an A5\,Vp\,SrEu star.

Abundances derived from high-resolution spectra show HD\,24355 to be slightly enhanced compared to some other roAp stars when considering the rare earth elements. However, a full and detailed abundance analysis is required to confirm its place amongst the roAp and noAp stars. The high-resolution spectra also allowed us to estimate a mean magnetic field strength of $2.64\pm0.49$\,kG; however, we take this to be an upper limit on the value due to the lack of Zeeman splitting in the spectra, and the method used to derive that value.

There is a discrepancy between the $T_{\rm eff}$ of HD\,24355 when using different methods to derive the parameter. Values from the literature, SED fitting abundance analysis and line fitting provide a wide range of $T_{\rm eff}$ values. However, we get the best agreement in results using solely the Balmer lines of both the low-resolution and high-resolution spectra, deriving $8200\pm200$\,K, placing HD\,24355 amongst the hotter roAp stars. 

Analysis of the pulsation mode as detected in the K2 data have shown the characteristic signatures of a roAp pulsator as predicted by the oblique pulsator model \citep{kurtz82,bigot02,bigot11}. The pulsational amplitude is modulated with the rotation period of the star, with the extremes in light variations and amplitude occurring at the same phase, indicative that the magnetic and pulsation poles lie in the same plane. The behaviour of the pulsation phase is not as expected, however. The very small phase change at quadrature is a surprise for a quadrupole pulsator. Examples from the literature (cf. Fig.\,\ref{fig:other_phases}) show a clear $\pi$-rad phase change when a different pulsation pole rotates into view, but this is not the case for HD\,24355. Here we see a shift of only 1-rad at most. This small `blip' in the phase suggests that HD\,24355 is pulsating in a very distorted mode, the most extreme case yet observed.

The rotationally split pulsation has provided us with the amplitudes to test the geometry of the star. As such, we modelled the system following the method of \citet{saio05}. Changing values of the inclination and obliquity angles and the polar magnetic field strength, and searching the parameter space surrounding our observational constraints, we conclude that HD\,24355 is a distorted quadrupolar pulsator, with a magnetic field strength of about 1.4\,kG. The model accurately matches the observed amplitude modulation of the pulsation, and the amplitudes of the rotationally split sidelobes. The pulsational phase variations are a stronger function of the evolution of the star, and as such provide a slightly greater challenge to model. We believe, however, that the model presented is a satisfactory match to the data, given our current observational constraints on the evolutionary stage of the star. We determine that the pulsation seen in HD\,24355 is super-critical, making it the most precisely observed super-critical roAp star to date. The driving mechanism for such a pulsation is currently unknown, thus making HD\,24355 a highly important target in understanding how some roAp stars can pulsate with frequencies well above the critical cutoff frequency. 

\section*{Acknowledgements}

DLH acknowledges financial support from the STFC via grant ST/M000877/1. GH is grateful for financial support from the Polish National Science Center, grant no. 2015/18/A/ST9/00578. The WASP project is funded and maintained by Queen's University Belfast, the Universities of Keele, St. Andrews and Leicester, the Open University, the Isaac Newton Group, the Instituto de Astrofisica Canarias, the South African Astronomical Observatory and by the STFC. This work made use of PyKE \citep{still12}, a software package for the reduction and analysis of Kepler data. This open source software project is developed and distributed by the NASA Kepler Guest Observer Office. We thank the anonymous referee for useful comments and suggestions.

\bibliography{24355refs}

\label{lastpage}
\end{document}